\documentclass[pdftex,prx,aps,nofootinbib,floatfix,superscriptaddress]{revtex4-2}
\usepackage[utf8]{inputenc}
\usepackage{braket}
\usepackage{amsmath}
\usepackage{makecell}
\usepackage{geometry}
\usepackage{amsfonts}
\usepackage{graphicx}
\usepackage[T1]{fontenc}
\usepackage{hyperref}
\usepackage{amsthm}
\usepackage{amssymb}
\newtheorem{definition}{Definition}[section]

\newtheorem{claim}{Claim}[section]
\newtheorem{theorem}{Theorem}[section]

\usepackage{vwcol}
\usepackage{tikz}
\usepackage{mathtools}
\usepackage{tasks}
\usepackage{appendix}
\usepackage{algorithm}
\usepackage{algpseudocode}
\usepackage{makecell}
\usepackage{wrapfig}

\begin{document}

\begin{abstract}

Port-based teleportation (PBT) is a form of quantum teleportation in which no corrective unitary is required on the part of the receiver. Two primary regimes exist - deterministic PBT in which teleportation is always successful, but is imperfect, and probabilistic PBT, in which teleportation succeeds with probability less than one, but teleportation is perfect upon a success. Two further regimes exist within each of these in which the resource state used for the teleportation is fixed to a maximally entangled state, or free to be optimised.

Recently, works resolved the long-standing problem of efficiently implementing port-based teleportation, tackling the two deterministic cases for qudits. Here, we provide algorithms in all four regimes for qubits. Emphasis is placed on the practicality of these algorithms, where we give polynomial improvements in the known gate complexity for PBT, as well as an exponential improvement in the required number of ancillas (albeit in separate protocols). 

Our approach to the implementation of the square-root measurement in PBT can be directly generalised to other highly symmetric state ensembles. For certain families of states, such a framework yields efficient algorithms in the case that the Petz recovery algorithm for the square-root measurement runs in exponential time.

\end{abstract}

\title{Efficient Algorithms for All Port-Based Teleportation Protocols}
\author{Adam Wills}
\author{Min-Hsiu Hsieh}
\affiliation{Hon Hai Research Instiute, Taipei}
\author{Sergii Strelchuk}
\affiliation{DAMTP, Centre for Mathematical Sciences, University of Cambridge, Cambridge CB30WA, UK}

\maketitle

\section{Introduction}

The 30-year old quantum teleportation protocol \cite{bennett1993teleporting} is a fundamental operation in quantum information theory in which an unknown quantum state may be reliably transmitted between spatially separated parties using only shared quantum entanglement and the transmission of classical information between them. The standard algorithm requires that the receiver, often referred to as Bob, performs a unitary recovery operation on the teleported state which depends on the outcome of a measurement performed by the sender, Alice, transmitted to Bob via the classical channel. Several questions immediately arise; among them is whether teleportation is possible without any corrective operation on the part of the receiver.

Building on the work of Knill, Laflamme and Milburn \cite{knill2001scheme}, Ishizaka and Hiroshima introduced port-based teleportation \cite{ishizaka2008asymptotic,ishizaka2009quantum} to answer this question. Alice and Bob, once again spatially separated, this time share a $2N$-qubit (or more generally qudit) entangled state, where they each hold $N$ such systems, called ports. Alice holds a further qubit (or qudit) which she wishes to teleport, and does so by performing a POVM on her $N+1$ systems. From the outcome of this POVM, she sends classical information to Bob telling him in which of his ports the teleported state may be found.

One quite striking, and useful, feature of such a protocol, is its unitary equivariance - Bob may perform any operation $E$ to every one of his ports before teleportation has even taken place, and ultimately obtain the teleported state acted on by $E$ after teleportation and the choice of the correct port. However, this power of the protocol is also, in some ways, a curse, since the no-programming theorem \cite{nielsen1997programmable} forbids such behaviour taking place faithfully and deterministically in systems of finite dimensions.

Therefore, in the physical case of finite $N$, either unit teleportation fidelity or unit success probability must be sacrificed. Ishizaka and Hiroshima thus define two primary regimes: deterministic port-based teleportation (dPBT) and probabilistic port-based teleportation (pPBT), where in the former, teleportation is always considered to have succeeded, but the fidelity (averaged over uniformly distributed input states) falls below 1, whereas in the latter, teleportation succeeds with probability (averaged over all uniformly distributed input states) below 1, but is perfect in the case of success. In the case of dPBT, Alice's POVM has $N$ outcomes, each of which correspond to a port in which Bob may find the teleported state. For pPBT, Alice's POVM has $N+1$ outcomes, where the additional outcome now corresponds to the failure of the protocol.

In \cite{ishizaka2009quantum}, Ishizaka and Hiroshima further introduce two regimes within dPBT and pPBT. First is the case of `maximally entangled resource state', where the state shared by Alice and Bob is fixed to $\ket{\psi^-}^{\otimes N}$, where $\ket{\psi^-} = \frac{\ket{01}-\ket{10}}{\sqrt{2}}$, and the `optimised resource state' case, where this $2N$-qubit state is free to take any form. Within these four regimes, Ishizaka and Hiroshima analytically determine the optimal POVMs (and resource states in the case of optimisable resource states) that maximise fidelity in the case of dPBT and success probability in the case of pPBT. They found different POVMs in each case, where for dPBT, the fidelity tends to 1 in the limit $N \to \infty$, and similarly for success probability in the case of pPBT. Interestingly, it was later shown in \cite{leditzky2022optimality} that the same measurement could be optimal in both dPBT regimes given the appropriate choice of optimal resource state. While this is surprising at first, it is possible owing to the fact that the convex optimisation problems relevant to port-based teleportation may have multiple optimising solutions. 

The measurement in question for dPBT is an example of a pretty good measurement (or square-root measurement)~\cite{eldar2001quantum,dalla2015optimality,hausladen1994pretty,holevo1978asymptotically,harrow2012many}. Quantum algorithms for general pretty good measurements based on the Petz recovery channel were presented in \cite{gilyen2022quantum}, but these algorithms are inefficient for the case in question\footnote{For the pretty good measurement relevant to dPBT, the average of the state ensemble has exponentially small non-zero eigenvalue, as we will go on to see. This leads to the algorithm of \cite{gilyen2022quantum} running inefficiently in the dPBT case.}. To obtain an efficient algorithm for the square-root measurement of dPBT, and also for pPBT, it is essential to exploit the symmetries of the state ensemble at hand.

Several applications of port-based teleportation are known, both practical and more theoretical. While the no-programming theorem forbids the existence of a faithful and deterministic universal programmable processor in finite dimensions, the unitary equivariance property of PBT allows it to be used as either a probabilistic, or approximate universal programmable processor even with finite $N$, in the cases of pPBT and dPBT respectively. Our new algorithms for pPBT therefore open up this former case for the first time. In addition, port-based teleportation has applications in instantaneous non-local quantum computation and position-based cryptography \cite{beigi2011simplified}, where the use of PBT allows for an exponential decrease in the amount of entanglement required to attack such a protocol. There are further implications for PBT in holography \cite{may2019quantum,may2022complexity} and channel discrimination \cite{pirandola2019fundamental}.

\subsection{Our Results}

Recently, works resolved the long-standing problem of the efficient implementation of port-based teleportation \cite{fei2023efficient,grinko2023gelfand} by showing how the POVM for dPBT could be implemented by leveraging a theory of `twisted Schur-Weyl duality' - see \cite{nguyen2023mixed} for another closely related work. In this work, we give efficient algorithms for all four regimes of port-based teleportation - both dPBT and pPBT, with maximally entangled and optimised resource states. Because the same POVM is optimal for both dPBT cases, it will suffice to consider only three cases. We focus on practicality in this work, going to some lengths to keep gate complexities and ancilla counts down. To this end, we provide a polynomial speedup in the known gate complexity for port-based teleportation, also aiming to minimise the amount of `practically difficult' operations used, such as amplitude amplification. Furthermore, although not in the same protocols, we show how each regime of port-based teleportation may be executed using only $\mathcal{O}(\log(N))$ qubits - an exponential improvement over all other methods.

Interestingly, all of our implementations will come from relatively simple manipulations within the spin coupling formalism \cite{jordan2009permutational,wills2023generalised}, a manifestation of the usual Schur-Weyl duality, as opposed to the twisted (or mixed) Schur-Weyl duality employed in other works implementing PBT \cite{fei2023efficient,grinko2023gelfand,nguyen2023mixed}. On the one hand, not using twisted Schur-Weyl duality restricts our algorithms to qubits, rather than qudits. However, on the other hand, the simplicity of the building blocks of our algorithms likely affords them greater generalisability. Implementing other pretty good measurements in the common case that the state ensemble has some high degree of symmetry may be possible to do efficiently using the framework presented herein, which is very useful in cases where the algorithm of \cite{gilyen2022quantum} runs inefficiently, as is true for dPBT.

Schur coupling features heavily throughout. There is a choice to be made within each one of our algorithms. To perform the Schur coupling, one may either use the BCH Schur transform of \cite{bacon2005quantum} or the spin coupling Schur transform of \cite{wills2023generalised}. The former runs faster, but uses $\mathcal{O}(N\log(N))$ qubits, whereas the latter is slower, but uses $\mathcal{O}(\log(N))$ qubits. The choice of which of these to use will reflect in the gate complexity and ancilla count of the final algorithm. The gate complexities and ancilla counts for our deterministic port-based teleportation are summarised in Table \ref{dPBTResults}, where we note again that the same POVM may be used in dPBT for maximally entangled resource state and optimal resource state.
\renewcommand{\arraystretch}{2}
\begin{table}[h]
    \centering
    \begin{tabular}{|c||c|c||c|c|}
         \hline Resource State&\multicolumn{2}{c||}{\makecell{Maximally Entangled}}&\multicolumn{2}{c|}{Optimised}\\\hline
         Schur Transform&BCH&Spin Coupling&BCH&Spin Coupling\\\hline\hline
         \makecell{Clifford + T\\Gate Complexity}&$\tilde{\mathcal{O}}(N^{3/2})\text{polylog}(\frac{1}{\epsilon})$&$\tilde{\mathcal{O}}(N^{7/2})\text{polylog}(\frac{1}{\epsilon})$&$\tilde{\mathcal{O}}(N^{3/2})\text{polylog}(\frac{1}{\epsilon})$&$\tilde{\mathcal{O}}(N^{7/2})\text{polylog}(\frac{1}{\epsilon})$\\\hline
         \makecell{Ancilla Count}&$\mathcal{O}(N\log(N))$&$\mathcal{O}(\log(N))$&$\mathcal{O}(N\log(N))$&$\mathcal{O}(\log(N))$\\\hline
         \makecell{Asymptotic (Average)\\Teleportation Fidelity}&\multicolumn{2}{c||}{$f \sim 1-\mathcal{O}\left(\frac{1}{N}\right)$}&\multicolumn{2}{c|}{$f \sim 1-\mathcal{O}\left(\frac{1}{N^2}\right)$}\\\hline
    \end{tabular}
    \caption{We present only one algorithm for dPBT in Section \ref{dPBTAlgo}, since there is only one POVM to perform. Depending on the choice of resource state used, the teleportation fidelity obtained will differ as shown. Also, one has the choice over whether to use the BCH Schur transform \cite{bacon2005quantum} or the spin coupling transform \cite{wills2023generalised} in the algorithm, which will effect the gate complexities and ancilla counts in the way shown. The symbol $\tilde{\mathcal{O}}$ hides polylogarithmic factors, and $\epsilon$ measures accuracy of the Naimark unitary implementing the POVM.}
    \label{dPBTResults}
\end{table}
\renewcommand{\arraystretch}{1}

For pPBT, we present two main algorithms, one for the POVM of the maximally entangled resource state case in Section \ref{pPBTMESAlgo} and one for the POVM of the optimised resource state case in Section \ref{pPBTOptAlgo}. The results for these algorithms are summarised in Table \ref{pPBTResults}.

\renewcommand{\arraystretch}{2}
\begin{table}[h]
    \centering
    \begin{tabular}{|c||c|c||c|c|}
         \hline Resource State&\multicolumn{2}{c||}{\makecell{Maximally Entangled}}&\multicolumn{2}{c|}{Optimised}\\\hline
         Schur Transform&BCH&Spin Coupling&BCH&Spin Coupling\\\hline\hline
         \makecell{Clifford + T\\Gate Complexity}&$\tilde{\mathcal{O}}(N)\text{polylog}(\frac{1}{\epsilon})$&$\tilde{\mathcal{O}}(N^3)\text{polylog}(\frac{1}{\epsilon})$&$\tilde{\mathcal{O}}(N^{3/2})\text{polylog}(\frac{1}{\epsilon})$&$\tilde{\mathcal{O}}(N^{7/2})\text{polylog}(\frac{1}{\epsilon})$\\\hline
         \makecell{Ancilla Count}&$\mathcal{O}(N\log(N))$&$\mathcal{O}(\log(N))$&$\mathcal{O}(N\log(N))$&$\mathcal{O}(\log(N))$\\\hline
         \makecell{Asymptotic (Average)\\Success Probability}&\multicolumn{2}{c||}{$p \sim 1-\mathcal{O}\left(\frac{1}{\sqrt{N}}\right)$}&\multicolumn{2}{c|}{$p \sim 1-\mathcal{O}\left(\frac{1}{N}\right)$}\\\hline
    \end{tabular}
    \caption{Our two main algorithms for pPBT, with maximally entangled resource state and optimised resource state, are presented in Sections \ref{pPBTMESAlgo} and \ref{pPBTOptAlgo}, respectively.}
    \label{pPBTResults}
\end{table}
\renewcommand{\arraystretch}{1}

One notes that with optimised resource state, the gate complexities are the same for pPBT as for dPBT. However, one can also see that there is an improvement by a factor of $\mathcal{O}(\sqrt{N})$ from dPBT to pPBT with maximally entangled resource state. This is because, with a careful treatment of the POVM operators, we show that it is possible to execute this algorithm using only a constant (five, to be precise) rounds of oblivious amplitude amplification, as opposed to the $\mathcal{O}(\sqrt{N})$ rounds of amplitude amplification used in the other cases.

Aside from the use of the usual Schur-Weyl duality versus twisted Schur-Weyl duality, our algorithms bear some similarities to the one for dPBT in \cite{fei2023efficient}, in which an extensive use of amplitude amplification is employed. However, the constant number of rounds of amplitude amplification in our algorithm for pPBT (with maximally entangled resource state), as well as the lower overall gate complexities, gives this protocol a greater chance of being practical for near-term devices. Another advantage of our algorithm in this domain is the single qubit used for the block-encoded unitary matrices throughout the paper, in contrast to the logarithmic number used in \cite{fei2023efficient}.

Finally, the use of a small constant number of rounds of amplitude amplification in the algorithm of Section \ref{pPBTMESAlgo} for pPBT with maximally entangled resource state leads one to the question of how well one can do when no amplitude amplification is used at all. To this end, we present an alternative algorithm for pPBT with maximally entangled resource state in Section \ref{alternativepPBT} that is intended to maximise potential practicality for near-term devices. No amplitude amplification is employed at all, and the gate complexities and ancilla counts are the same for this algorithm as for our main algorithm for pPBT with maximally entangled resource state shown on the left of Table \ref{pPBTResults}. However, unfortunately, its overall success probability is $\frac{1}{4}$ of what it is for normal pPBT. As a result, the asymptotic success probability is $\frac{1}{4}$, rather than the usual 1. Nevertheless, in a practical situation, only a (small) constant number of copies of the state to be teleported would be required to overcome this. This algorithm therefore seems to form the best currently available option for a practical implementation of port-based teleportation.

\subsection{Concurrent Work}

Concurrently and independently of the work presented in this manuscript, Grinko, Burchardt and Ozols derived algorithms for all port-based teleportation protocols on qudits with $\text{poly}(n,d,\log 1/\epsilon)$ bounds on the gate complexity and ancilla count. Subsequent to discussions with these authors on our algorithms, they were able to adapt these ideas to their setting to derive similar complexity improvements, hence the temporal separation in the publications. By using phase estimation techniques (where we use amplitude amplification), they are able to derive further polynomial improvements in the gate complexity, as may be found in \cite{grinko2023efficient}.

\section{Outline of Results}

In Section \ref{prelimSection}, we give the necessary preliminaries on the Schur transform defined via $SU(2)$ spin coupling and a formal definition of port-based teleportation, in particular the POVMs used in each regime. Those for whom the spin coupling definition of the Schur transform is unfamiliar, but the more traditional definition as in the work of Harrow \cite{harrow2005applications} is familiar, are recommended to read Appendix \ref{preMapping}. This Appendix gives a translation between these two perspectives on the Schur transform, while also clearing up confusion that arises between the two definitions.

In Section \ref{diagonalise}, we perform the necessary analysis on the set of POVM operators in each case, in particular finding their eigendecomposition, which will be essential for our later implementations. Then, in Section \ref{algos}, we provide our algorithms for port-based teleportation. This Section is intended to be independent of Section \ref{diagonalise} - all necessary results proven in Section \ref{diagonalise} are repeated in Section \ref{algos}. The algorithm for dPBT may be found in Section \ref{dPBTAlgo}, the main algorithm for pPBT with maximally entangled resource state may be found in Section \ref{pPBTMESAlgo}, the alternative algorithm (using no amplitude amplification) for pPBT with maximally entangled resource state may be found in Section \ref{alternativepPBT} and lastly the algorithm for pPBT with optimised resource state may be found in Section \ref{pPBTOptAlgo}. A gate complexity analysis of these algorithms takes place in Appendix \ref{complexityAnalysis}, and lastly we provide a discussion in Section \ref{discussion}.

\section{Preliminaries}\label{prelimSection}

Since the notion will be used throughout this paper, we start by introducing $SU(2)$ spin couplings in Section \ref{spinPrelim} before defining the Schur transform in \ref{schurPrelim} and describing how it may be efficiently implemented via spin coupling. Lastly, in Section \ref{PBTPrelim}, we formally define port-based teleportation. 

Those who have prior experience with the Schur transform not from the spin coupling perspective, but from the more mathematical perspective such as that used by Harrow \cite{harrow2005applications} (the more common perspective in quantum computing) are recommended to read Appendix \ref{preMapping} for the translation between the two perspectives. 

\subsection{Spin Couplings}\label{spinPrelim}

A coupling of $n$ spins is described via angular momenta, defined as follows.

\begin{definition}
    On $n$ qubits, we let $(\sigma_x^{(i)},\sigma_y^{(i)},\sigma_z^{(i)})$ denote the usual Pauli matrices acting on the $i$-th qubit. The angular momentum operator on the $i$-th qubit is then defined via
    \begin{equation}
        \vec{S}^{(i)} = \frac{1}{2}\begin{pmatrix} \sigma_x^{(i)} \\ \sigma_y^{(i)} \\ \sigma_z^{(i)} \end{pmatrix}.
    \end{equation}
\end{definition}
It will then be important to be able to express so-called $Z$-angular momentum and total angular momentum as operators acting on a subset of the qubits.

\begin{definition}
    Given a subset of $n$ qubits denoted by the subset $a \subseteq [n] = \{1, ..., n\}$, their total angular momentum operator is
    \begin{equation}
        S^2_a = \left(\sum_{i \in a}\vec{S}^{(i)}\right) \cdot \left(\sum_{i \in a} \vec{S}^{(i)}\right),
    \end{equation}
    where $\cdot$ denotes the usual dot product of 3-vectors, and their $Z$-angular momentum is
    \begin{equation}
        Z_a = \frac{1}{2}\sum_{i \in a}\sigma_z^{(i)}.
    \end{equation}
\end{definition}

If a subset of qubits $a$ is in a state that is an eigenvector of $S^2_a$ with eigenvalue $j(j+1)$, the set of qubits is said to have total spin $j$ (or simply just ``spin $j$''). It can be verified that $S^2_{\{i\}} = \frac{3}{4}I$ for an individual qubit $i$, so we treat individual qubits as having spin $1/2$. If a subset of qubits $a$ is in a state that is an eigenvector of $Z_a$ with eigenvalue $m$, the set of qubits is said to have $Z$-spin $m$.

The following fact is important for us and has a simple proof \cite{SchurSampling}:

\begin{claim}
    For non-empty subsets of the qubits $a$ and $b$, we have the following commutativity between operators:
    \begin{align}
        a \cap b = \emptyset &\implies [S^2_a,S^2_b] = 0\\
        a \subseteq b &\implies [S^2_a,S^2_b] = 0\\
        a \subseteq b &\implies [S^2_a,Z_b] = 0.
    \end{align}
\label{commutativity}\end{claim}

In particular, we see that $[S^2_{[n]},Z_{[n]}] = 0$. We can thus write out a basis for the space of the $n$ qubits where each basis state has total spin $j$ and $Z$-spin $m$, for some $j$ and $m$, denoting such a simultaneous eigenstate as $\ket{j,m}$. $j$ only takes values in $\frac{\mathbb{N}_0}{2}$ for any set of qubits, and given some fixed $j$, $m$ only takes values in $\{-j, -j+1, ..., j-1, j\}$. However, $\ket{j,m}$ does not identify one particular state of $n$ qubits in general. For example, it can be checked that the dimension of the eigenspace with $j=1/2$ and $m=1/2$ on 3 qubits is greater than 1. In order to fix this degeneracy, and uniquely identify states via their various spins, we need the concept of a spin eigenbasis.

A spin eigenbasis is a simultaneous eigenbasis arising from a complete set of $n$ commuting total and $Z$-angular momentum operators. The allowed choices of such operators are explained in the following. 

\begin{definition}
    Consider non-empty, proper qubit subsets $a_1, ..., a_{n-2} \subseteq [n]$ such that, for each $a_i$ and $a_j$, they are either disjoint or one is a subset of the other. A spin eigenbasis is then the simultaneous eigenbasis corresponding to the set of commuting observables $(S^2_{a_1}, ..., S^2_{a_{n-2}},S^2,Z)$ where we adopt the short-hand $S^2 \coloneqq S^2_{[n]}$, $Z \coloneqq Z_{[n]}$.
\end{definition}
Such sets of operators are indeed commuting by Claim \ref{commutativity}, and complete \cite{jordan2009permutational}. As an example, on three qubits, we can identify a basis via the choice of complete and commuting operators $(S^2_{\{1,2\}},S^2,Z)$. We have thus fixed the previous degeneracy by using a complete set of operators, and states in this basis are uniquely identified as $\ket{k,j,m}$, where $k$ is now the spin on the first two qubits. 

To get a better grasp of these bases, we introduce the notion of spin coupling. Spin coupling is defined via Clebsch-Gordan coefficients, where here we deal only with $SU(2)$ Clebsch-Gordan coefficients, denoted $C^{J,M}_{j_1, m_1; j_2, m_2}$. Given some set of qubits $a$ with total spin $j_1$, whose $Z$-spin $m_1$ is allowed to vary, and similarly some other set of qubits $b$ with total spin $j_2$, whose $Z$-spin $m_2$ is allowed to vary, the two sets of qubits may be brought together to form a state on qubits $a \cup b$ which has definite total spin $J$ and $Z$-spin $M$:

\begin{equation}
    \ket{J,M} = \sum_{m_1, m_2}C^{J,M}_{j_1, m_1; j_2, m_2}\ket{j_1, m_1}\ket{j_2, m_2}\label{CGCoupling}
\end{equation}
where the sum is taken over the allowed values of $m_1$ and $m_2$ mentioned earlier, i.e. $m_1 \in \{-j_1, ..., j_1-1, j_1\}$ etc. The state in Equation \ref{CGCoupling} is an eigenstate of $S^2_{a \cap b}$, $Z_{a \cap b}$, $S^2_a$ and $S^2_b$, but it is not an eigenstate of either $Z_a$ or $Z_b$, like $\ket{j_1, m_1}$ and $\ket{j_2, m_2}$ are, respectively. Given some $j_1$ and $j_2$ both in $\frac{\mathbb{N}_0}{2}$, which total spins $J$ can we get? Any $\ket{J,M}$ can be produced with $M \in \{-J, ..., J-1, J\}$ as long as

\begin{equation}
    |j_1 - j_2| \leq J \leq j_1 + j_2 \;\;\text{ and }\;\; j_1 + j_2 + J \in \mathbb{Z}\label{AMRules}
\end{equation}
and so, in particular, $n$ qubits may have any total spin in $\{\frac{1}{2}, \frac{3}{2}, ..., \frac{n}{2}\}$ if $n$ is odd, or in $\{0, 1, ..., \frac{n}{2}\}$ if $n$ is even. If $j_1$, $j_2$ and $J$ do not satisfy the relations of Equation \eqref{AMRules}, any corresponding Clebsch-Gordan coefficient will return the value $0$. Another important property of Clebsch-Gordan coefficients is that $C^{J,M}_{j_1, m_1; j_2, m_2}= 0$ unless $m_1 + m_2 = M$, a property referred to as ``conservation of angular momentum''.

The Clebsch-Gordan coefficients $C^{J,M}_{j_1, m_1; j_2, m_2}$ are numbers that can be calculated classically in \\$\text{poly}(J,M,j_1, m_1, j_2, m_2)$ time \cite{enwiki:1182363913,Wigner3jWolfram}. From here, we see how we can build up spin eigenbases. As an example, we can find the states given by coupling two qubits:

\begin{align}
    \ket{j=1,m=-1} = \ket{11} \hspace*{1cm} \ket{j=1, m=0} &= \frac{\ket{01}+\ket{10}}{\sqrt{2}} \hspace*{1cm} \ket{j=1, m=1} = \ket{00} \\
    \ket{j=0,m=0} &= \frac{\ket{01}-\ket{10}}{\sqrt{2}}
\end{align}
by identifying the computational basis states of a qubit $\ket{0}$ and $\ket{1}$ with spin states $\ket{j=\frac{1}{2}, m = \frac{1}{2}}, \ket{j=\frac{1}{2}, m = -\frac{1}{2}}$, respectively. We then further couple these to get the 8 states in the spin eigenbasis on 3 qubits defined by $(S^2_{1,2},S^2,Z)$. This eigenbasis then contains states $\ket{k,j,m}$ as follows:

\begin{align}
    \Ket{1, \frac{3}{2}, -\frac{3}{2}} \hspace*{1cm} \Ket{1, \frac{3}{2}, -\frac{1}{2}}\hspace*{1cm}&\Ket{1, \frac{3}{2}, \frac{1}{2}}\hspace*{1cm} \Ket{1, \frac{3}{2}, \frac{3}{2}} \\
    \Ket{1, \frac{1}{2}, -\frac{1}{2}}\hspace*{1cm}&\Ket{1, \frac{1}{2}, \frac{1}{2}} \\
    \Ket{0, \frac{1}{2}, -\frac{1}{2}}\hspace*{1cm}&\Ket{0, \frac{1}{2}, \frac{1}{2}}
\end{align}

\subsection{The Schur Transform}\label{schurPrelim}

Here we define the Schur transform from the physical point of view, via spin coupling. Those for whom this is a new perspective may find Appendix \ref{preMapping} useful in which we discuss the two different notions of the Schur transform and why it's important to appreciate both of them. We also provide a translation between the notations in the mathematical and physical perspectives to ease this transition.

For us, the Schur basis is a spin eigenbasis of particular importance.

\begin{definition}
    The Schur basis on $n$ qubits is the spin eigenbasis defined by the operators \\$\left(S^2_{\{1,2\}}, S^2_{\{1,2,3\}}, ..., S^2_{\{1, ..., n-1\}}, S^2, Z\right)$. The Schur transform is the unitary operator on $n$ qubits mapping the computational basis state to the Schur basis.
\end{definition}

States in this basis are called Schur states and may then be labelled by $n-1$ total spins and one $Z$-spin. We write states in this basis as $\ket{k_1, k_2, ..., k_{n-2}, j, m}$, where $k_1$ is the spin on the first two qubits, $k_2$ is the spin on the first three qubits, and so on. $k_{n-2}$ is therefore the spin on the first $n-1$ qubits, and $j$ is the spin on all the qubits. It is readily verified that these may be explicitly written

\begin{multline}
    \ket{k_1, k_2, ..., k_{n-2},j,m} = \sum_{x_1, ..., x_n}C^{k_1, x_1 + x_2}_{\frac{1}{2},x_1;\frac{1}{2},x_2}\;C^{k_2, x_1 + x_2 + x_3}_{k_2, x_1 + x_2; \frac{1}{2},x_3}\; ... \\ C^{k_{n-2},x_1 + ... + x_{n-1}}_{k_{n-3},x_1 + ... + x_{n-2};\frac{1}{2},x_{n-1}}\;C^{j,m}_{k_{n-2},x_1+...+x_{n-1};\frac{1}{2},x_n}\ket{x_1 ... x_n}\label{schurState}
\end{multline}

For convenience, in this expression, each $x_i$ is being summed over the set $\{\pm \frac{1}{2}\}$ and we are writing out computational basis states $\ket{x_i}$ with a slightly different notation to usual: $\Ket{-\frac{1}{2}}$ instead of $\ket{1}$ and $\Ket{\frac{1}{2}}$ instead of $\ket{0}$.

We note that the definition given above is a definition from a very physical perspective, inspired by the work on Permutational Quantum Computing from Jordan \cite{jordan2009permutational} which dates back to the work of Marzuoli and Rasetti \cite{marzuoli2005computing}. Much of the more modern literature approaches the Schur transform from a more mathematical, but essentially equivalent perspective in relation to the beautiful theory of Schur-Weyl duality. See \cite{bacon2005quantum} for an excellent introduction. We note that in this more mathematical formalism all spin eigenbases are examples of Schur bases, but \textit{the} Schur basis given above is one that is subgroup-adapted in a particular way.

Algorithms exist for efficiently performing the Schur transform on a quantum computer. See \cite{wills2023generalised} for an algorithm that fits into this perspective. Briefly, this algorithm takes place in two stages. First, computational basis states are coherently mapped to ``encodings'' of Schur states via the ``pre-mapping'' stage:

\begin{equation}
    \ket{x_1...x_n} \mapsto \ket{k_1}\ket{k_2}...\ket{k_{n-2}}\ket{j}\ket{m}.\label{preMapEqn}
\end{equation}
Each ket on the right-hand side is a computational basis state encoding its displayed value. For example, $k_3$ may take three values 0, 1 and 2, and so $\ket{k_3}$ is a computational basis state on two qubits. In total, there are $\mathcal{O}(n\log(n))$ qubits on the right-hand side. Next, these Schur encodings are mapped to Schur states themselves:

\begin{equation}
    \ket{k_1}\ket{k_2}...\ket{k_{n-2}}\ket{j}\ket{m} \mapsto \ket{k_1, k_2, ..., k_{n-2},j,m}\label{couplingEqn}
\end{equation}
in what is known as the ``coupling'' stage. This again takes place coherently over all Schur encodings. Notice that the right-hand side is a state on $n$ qubits now - the ancillas have become unentangled and discarded. Throughout, we will differentiate Schur states from Schur encodings by using one ket $\ket{k_1, k_2, ..., k_{n-2}, j, m}$ and multiple kets $\ket{k_1}\ket{k_2}...\ket{k_{n-2}}\ket{j}\ket{m}$, respectively.

The Schur transform has found wide-spread use in quantum information applications. One point that is often under-appreciated but critical in some of these applications is the need to perform a ``clean'' transform, where $n$-qubit computational basis states $\ket{x}$ are mapped to $n$-qubit Schur states $\ket{k_1, ..., k_{n-2},j,m}$ and vice-versa. This is important whenever further unitary $n$-qubit computation will take place after the Schur transform, such as in \cite{jordan2009permutational} or \cite{zheng2022super}. For other applications, often, it is only the inverse of the coupling stage that is used, where $n$ qubit Schur states are mapped to their Schur encodings. In fact, this operation itself is defined in many places as the Schur transform, and there are efficient implementations of this \cite{harrow2005applications,kirby2017practical,krovi2019efficient}, but this has led to confusion in some places for algorithm designers who require a clean transform\footnote{We note that \cite{krovi2019efficient} can also perform a clean transform. However, the algorithm of \cite{wills2023generalised} offers the lowest gate complexity in terms of the Clifford + T universal set of any clean transform, as well as the greatest simplicity, and the ability to perform the unitary to any spin eigenbasis, not just the Schur transform.}. In Appendix \ref{preMapping}, we go into further detail on the discrepancies between different notions of the Schur transform, addressing particular sticking points where confusion can arise.

Finally, we note that the above algorithm, as outlined in Equations \eqref{preMapEqn} and \eqref{couplingEqn}, uses $\mathcal{O}(n\log(n))$ ancillas. However, a ``compressed'' version is shown in \cite{wills2023generalised} that uses only $\mathcal{O}(\log(n))$ ancillas. In this version the pre-mapping stage is
\begin{equation}
    \ket{x_1...x_n} \mapsto \ket{y_1...y_{n-2}}\ket{k_{n-2}}\ket{j}\ket{m},
\end{equation}
where $\ket{y_1...y_{n-2}}$ is an $(n-2)$-qubit computational basis state giving an efficient encoding of the values $k_1, ..., k_{n-2}$, and then the coupling stage performs
\begin{equation}
    \ket{y_1...y_{n-2}}\ket{k_{n-2}}\ket{j}\ket{m} \mapsto \ket{k_1, k_2, ..., k_{n-2},j,m}.
\end{equation}
The Clifford + T gate complexity of this is $\mathcal{O}(n^3\log(n)\log(\frac{n}{\epsilon}))$, where $\epsilon$ is its accuracy in the trace norm.

In this work, all that will be needed will be the inverse of the pre-mapping stage, which for many people is just the Schur transform, as mentioned. It will also be important for us that the algorithm of \cite{bacon2005quantum} gives an algorithm for this operation with gate complexity $n\;\text{poly}(\log n,\log\frac{1}{\epsilon})$, but using $\mathcal{O}(n\log(n))$ ancillas.

\subsection{Port-Based Teleportation}\label{PBTPrelim}

Ishizaka and Hiroshima introduced the `port-based' method of quantum teleportation in \cite{ishizaka2008asymptotic,ishizaka2009quantum} based on the work in \cite{knill2001scheme}. The scheme aims to offer a method of teleportation for which no corrective unitary operation is required on the part of the receiver, in contrast to the standard teleportation protocol \cite{bennett1993teleporting}. It may be described as follows, in a similar way to \cite{ishizaka2009quantum}.

The sender, Alice, and receiver, Bob, assumed to be spatially separated, each have $N$ qubit systems, respectively labelled $A_1, ..., A_N$ and $B_1, ..., B_N$ in a joint entangled state. The state Alice wishes to send is in a further qubit system which we denote $A_{N+1}$. She performs a POVM on her $N+1$ qubit systems and sends her result, a number in $\{1, ..., N\}$ to Bob as classical information. This number tells Bob in which port he may find the teleported state, and he need do nothing other than select that port.

One of the most attractive features of such a scheme is its `unitary equivariance', meaning that Bob can perform some unitary $U$ on every port before the classical information has even arrived telling him which port to look in, ultimately obtaining the teleported state having been acted on by $U$. Unfortunately, the no-programming theorem \cite{nielsen1997programmable} prohibits this process from taking place perfectly (with fidelity 1) and deterministically (with probability 1) in a system of finite size. Thus, with finite $N$, one must sacrifice at least one of the ability to perform the teleportation perfectly or deterministically.

Two primary paradigms then emerge, that of deterministic port-based teleportation (dPBT) and probabilistic port-based teleportation (pPBT). In the former, teleportation is always achieved, but with fidelity less than 1 (strictly, this is fidelity averaged over all uniformly distributed pure input states), and in the latter, teleportation is achieved with probability less than 1 (again, strictly, this is averaged probability) but fidelity 1 in the case of a success. Note that in this latter case, Alice in fact has $N+1$ outcomes from her measurement, the first $N$ denoting a success (and the corresponding port in which to find the state) and the final one, $N+1$, denoting a failure. In both regimes, Ishizaka and Hiroshima determine the POVM that will optimise fidelity and probability respectively. The optimal fidelity and probability tend towards 1 as $N \to \infty$ in each case.

Furthermore, Ishizaka and Hiroshima define two more regimes within dPBT and pPBT: one with a maximally entangled resource state, where the shared resource state between the $2N$ ports is fixed to $\ket{\psi^-}^{\otimes N}$, where $\ket{\psi^-} = \frac{\ket{01}-\ket{10}}{\sqrt{2}}$, and one for which this resource state is simultaneously optimised with the POVM.
 
For the four regimes defined, Ishizaka and Hiroshima find the following optimal POVMs. For dPBT with maximally entangled resource state, the optimal POVM operators are given by the square-root measurement, also called the pretty good measurement

\begin{equation}
    \Pi_i = \rho^{-1/2}\sigma^{(i)}\rho^{-1/2} + \Delta \;\text{ for } \;i = 1, ..., N
\end{equation}
where $\sigma^{(i)} = \frac{1}{2^{N-1}}\ket{\psi^-}\bra{\psi^-}_{A_iA_{N+1}}\otimes I_{\bar{A}_i}$ for $\bar{A}_i = A_1...A_{i-1}A_{i+1}...A_N$ and $\rho = \sum_{i=1}^N\sigma^{(i)}$. The inverse $\rho^{-1}$ is defined only on the support of $\rho$ and the term

\begin{equation}
    \Delta = \frac{1}{N}\left(I - \sum_{i=1}^N\rho^{-1/2}\sigma^{(i)}\rho^{-1/2}\right)
\end{equation}
is added to every POVM element so that $\sum_{i=1}^N \Pi_i = I$. Ishizaka and Hiroshima then give a separate measurement for the case of dPBT with optimised resource state, however it was shown in \cite{leditzky2022optimality} that the pretty good measurement is also optimal in this case too - note this may be the case because the relevant optimisation problem may have multiple solutions.

For pPBT with maximally entangled resource state, Ishizaka and Hiroshima show that the success probability is optimised by the POVM operators

\begin{equation}
    \Pi_i = \ket{\psi^-}\bra{\psi^-}_{A_iA_{N+1}}\otimes\tilde{\Theta}_{i\bar{A_i}}\; \text{ for } \;i = 1, ..., N \;\text{ and }\; \Pi_{N+1} = I - \sum_{i=1}^N\Pi_i
\end{equation}
where $\tilde{\Theta}_{i\bar{A}_i}$ denotes the operator $\tilde{\Theta}_i$ acting on the qubits $\bar{A}_i = A_1...A_{i-1}A_{i+1}...A_N$. This is an $N-1$ qubit operator defined as

\begin{equation}
    \tilde{\Theta}_i = \frac{1}{2^{N-1}}\sum_{s=s_{min}}^{(N-1)/2}\frac{1}{\lambda^+_{s+1/2}}\mathbb{I}(s)
\end{equation}
where

\begin{equation}
    s_{min} = \begin{cases} 0 & \text{ if } N-1 \text{ is even} \\ \frac{1}{2} & \text{ if } N-1 \text{ is odd} \end{cases}, \;\;\;\;\;\;\; \lambda^+_j = \frac{1}{2^N}\left(\frac{N}{2}+j+1\right)
\end{equation}
and $\mathbb{I}(s)$ is the projector onto the subspace of the $N-1$ qubits with total spin $s$. Lastly, for pPBT with optimised resource state, Ishizaka and Hiroshima show that optimal POVM operators are given by

\begin{equation}
    \Pi_i = (O^{-1}_A\otimes I_{A_{N+1}})\left(\ket{\psi^-}\bra{\psi^-}_{A_iA_{N+1}}\otimes \tilde{\Theta}_{i\bar{A}_i}\right)(O^{-1}_A\otimes I_{A_{N+1}})\; \text{ for } \;i = 1, ..., N \;\text{ and }\; \Pi_{N+1} = I - \sum_{i=1}^N\Pi_i
\end{equation}
where $A = A_1...A_N$ denotes Alice's ports and
\begin{align}
    O &= \sum_{j = j_{min}}^{N/2}\sqrt{\nu(j)}\mathbb{I}(j)_A \;\;\;\text{ for }\;\;\; \nu(j) = \frac{2^Nh(N)(2j+1)}{g^{[N]}(j)} \;\;\;\text{ and }\;\;\; j_{min} = \begin{cases}0 & \text{ if } N \text{ is even} \\ \frac{1}{2} & \text{ if } N \text{ is odd}\end{cases}\\
    \tilde{\Theta}_i &= \sum_{s=s_{min}}^{(N-1)/2}u(s)\mathbb{I}(s)\;\;\;\;\;\;\;\text{ for }\;\;\;u(s) = \frac{2^{N+1}h(N)(2s+1)}{Ng^{[N-1]}(s)}
\end{align}
where $\mathbb{I}(j)$ is the projector onto the subspace of $N$ qubits with spin $j$, $h(N) = \frac{6}{(N+1)(N+2)(N+3)}$ and $g^{[N]}(j) = \frac{(2j+1)N!}{(\frac{N}{2}-j)!(\frac{N}{2}+1+j)!}$.

Ishizaka and Hiroshima show that the optimal fidelities (for dPBT) and success probabilities (for pPBT) behave as in Table \ref{optimalBehaviour}.
\renewcommand{\arraystretch}{2}
\begin{table}[]
    \centering
    \begin{tabular}{c||c|c}
    & \makecell{Optimal\\Fidelity (dPBT)} & \makecell{Optimal\\Success Probability (pPBT)}\\\hline\hline
         \makecell{Maximally Entangled\\Resource State}& $f \sim 1 - \mathcal{O}\left(\frac{1}{N}\right)$& $p \sim 1-\mathcal{O}\left(\frac{1}{\sqrt{N}}\right)$\\\hline
         \makecell{Optimised\\Resource State}&$f \sim 1-\mathcal{O}\left(\frac{1}{N^2}\right)$ & $p \sim 1-\mathcal{O}\left(\frac{1}{N}\right)$
    \end{tabular}
    \caption{Optimal fidelities and success probabilities for the four regimes in the limit $N \to \infty$. Notice that for both dPBT and pPBT we get a quadratic improvement in the `deviation from 1' by moving to the optimised protocol.}
    \label{optimalBehaviour}
\end{table}
\renewcommand{\arraystretch}{1}

\section{Diagonalising the PBT Operators}\label{diagonalise}

Noting that we can use the same POVM operators for both cases in dPBT, we now study the three sets of POVM operators - those for dPBT (Section \ref{dPBTPOVM}), those for pPBT with maximally entangled resource state (Section \ref{pPBTMESPOVM}) and those for pPBT with optimised resource state (Section \ref{pPBTOptiPOVM}).

Our aim will be to diagonalise these POVM operators. We will find that it is relatively straightforward to do so for all these POVMs using just standard spin coupling, demonstrating the power of the formalism. The diagonalisation is necessary so that we can perform the POVM via Naimark dilation. Indeed, a standard way to perform a POVM described by the operators $\{\Pi_i\}_{i=1}^M$ on the state $\ket{\psi}$, is to attach an ancillary register $\ket{i}$ of dimension $M$ and perform the operation

\begin{equation}
    \ket{\psi}\ket{1} \mapsto \sum_{i=1}^M\sqrt{\Pi_i}\ket{\psi}\ket{i}\label{naimarkUnitary}
\end{equation}
where $\{\ket{i}\}_{i=1}^M$ forms an orthonormal basis for the ancillary register. Then, by measuring the ancillary register projectively, one recovers the outcome $i$ with the desired probability $\bra{\psi}\Pi_i\ket{\psi}$. Note that the operation in Equation \eqref{naimarkUnitary} is in general quite non-trivial to implement, but because the operation on the right-hand side has norm 1, we are guaranteed that there is some unitary acting in this way - a so-called ``Naimark unitary''.

To diagonalise the POVM operators, we will need a consistent notation for Schur states on $N+1$ qubits. We denote a Schur state on $N+1$ qubits by $\ket{k_1, ..., k_{N-2}, j, s, m}$, where $k_1$ is the spin on the first two qubits, $k_2$ is the spin on the first three qubits, and so on. $k_{N-2}$ is therefore the spin on the first $N-1$ qubits, which in our case are the qubits $A_1, ..., A_{N-1}$, and $j$ is the spin on the first $N$ qubits, i.e. $A = A_1, ..., A_N$, all of Alice's ports. $s$ is then the spin on all $N+1$ qubits (all of Alice's ports and the state she intends to send), and $m$ is the $Z$-spin on all these qubits\footnote{We note that Appendix \ref{preMapping} contains a useful translation between this physical (spin) perspective of the Schur states and the more common mathematical perspective which will likely be of use to those who are previously familiar with the latter perspective only.}.

Often, this notation will prove too cumbersome, and we in fact find it better to simply write $\ket{k,j,s,m}$ where $k$ is standing for all $k_1, ..., k_{N-2}$. Sometimes, we will want to address $k_{N-2}$, but none of the other $k$'s. In this case, we will write the Schur state as $\ket{\tilde{k},k_{N-2},j,s,m}$, where $\tilde{k}$ is standing for $k_1, ..., k_{N-3}$.

\subsection{Deterministic Port-Based Teleportation}\label{dPBTPOVM}

Let us study the POVM operators used in dPBT, i.e. the square-root measurement. Acting on Alice's ports $A = A_1...A_N$ and the input state $A_{N+1}$, we recall that these are $\Pi_i = \rho^{-1/2}\sigma^{(i)}\rho^{-1/2} + \Delta \;\text{ for } \;i = 1, ..., N $ where $\sigma^{(i)} = \frac{1}{2^{N-1}}\ket{\psi^-}\bra{\psi^-}_{A_iA_{N+1}}\otimes I_{\bar{A}_i}$ for $\bar{A}_i = A_1...A_{i-1}A_{i+1}...A_N$, $\rho = \sum_{i=1}^N\sigma^{(i)}$ and $\ket{\psi^-} = \frac{\ket{01}-\ket{10}}{\sqrt{2}}$. The inverse $\rho^{-1}$ is defined only on the support of $\rho$ and the term

\begin{equation}
    \Delta = \frac{1}{N}\left(I - \sum_{i=1}^N\rho^{-1/2}\sigma^{(i)}\rho^{-1/2}\right)
\end{equation}
is added to every POVM element so that $\sum_{i=1}^N \Pi_i = I$.

Ishizaka and Hiroshima show that the operator $\rho$ is diagonalised by the Schur basis $\ket{k,j,s,m}$, and in particular

\begin{equation}
    \rho\ket{k,j,s,m} = \lambda(j,s)\ket{k,j,s,m} \;\text{ where } \;\lambda(j,s) = \begin{cases} \lambda^-_j = \frac{1}{2^N}\left(\frac{N}{2}-j\right) & \text{ if } s = j + \frac{1}{2} \\ \lambda^+_j = \frac{1}{2^N}\left(\frac{N}{2} + j + 1\right) & \text{ if } s = j - \frac{1}{2}\end{cases}.
\end{equation}
From here, it is clear to see where $\rho$ is and is not supported. We see that $\rho$ has zero eigenvalue exactly when $j = \frac{N}{2}$ and $s = j + \frac{1}{2}$ which are, in fact, the maximal spin states on the $N+1$ qubits - these are the states with $k_1 = 1$, $k_2 = \frac{3}{2}$, $k_3 = 2$ and so on, up to $j = \frac{N}{2}$ and $s = \frac{N+1}{2}$. There are, in fact, $N+2$ such states, where $m$ is the only spin allowed to vary.

From here, we see that $\Delta = \frac{1}{N}\left(I - \rho^{-1/2}\rho\rho^{-1/2}\right)$, recalling that $\rho^{-1}$ is the inversion of $\rho$ only on its support. Therefore,

\begin{equation}
    \Delta = \frac{1}{N}\mathbb{I}\left(\frac{N+1}{2}\right)_{AA_{N+1}}
\end{equation}
where $\mathbb{I}\left(\frac{N+1}{2}\right)_{AA_{N+1}}$ is the projector onto the maximal spin states (with spin $\frac{N+1}{2}$) of all $N+1$ qubits. We therefore see that $\Delta$ simply acts to attach a factor of $\frac{1}{N}$ to each maximal spin state, whereas $\rho^{-1/2}\sigma^{(i)}\rho^{-1/2}$ annihilates every maximal spin state. $\Delta$ annihilates any Schur state with non-maximal spin, so it remains to find how $\rho^{-1/2}\sigma^{(i)}\rho^{-1/2}$ acts on Schur states of non-maximal spin. We will in fact just determine the action of $\Pi_N$, and then use $\Pi_i = SWAP_{A_iA_N}\Pi_NSWAP_{A_iA_N}$.

By using the expression for $(N+1)$-qubit Schur states in Equation \eqref{schurState}, as well as the explicit formulae for Clebsch-Gordan coefficients \cite{enwiki:1182363913}

\begin{equation}
    C^{\;j \pm \frac{1}{2},M}_{j,\;M-\frac{1}{2}\;;\frac{1}{2},\frac{1}{2}} = \pm \sqrt{\frac{1}{2}\left(1 \pm \frac{M}{j + \frac{1}{2}}\right)} \;\text{ and }\;
    C^{\;j \pm \frac{1}{2},M}_{j,\;M+\frac{1}{2}\;;-\frac{1}{2},\frac{1}{2}} = \sqrt{\frac{1}{2}\left(1 \mp \frac{M}{j+\frac{1}{2}}\right)},
\end{equation}
one finds the partial inner product

\begin{equation}
    \bra{\psi^-}_{A_NA_{N+1}}\ket{k,j,s,m} = c(k_{N-2},j,s)\ket{k_1, ..., k_{N-2},m}_{\bar{A}_N}
\end{equation}
where, to be explicit, $\ket{\psi^-}_{A_NA_{N+1}}$ is the singlet state, $\frac{\ket{01}-\ket{10}}{\sqrt{2}}$, on the last two qubits $A_N$ and $A_{N+1}$, $\ket{k,j,s,m}$ is an $(N+1)$-qubit Schur state. $\ket{k_1, ..., k_{N-2},m}_{\bar{A}_N}$ is an $(N-1)$-qubit Schur state on the qubits $\bar{A}_N = A_1...A_{N-1}$, with spin on the first two qubits $k_1$, on the first three qubits $k_2$, up to spin on all $N-1$ qubits of $k_{N-2}$ and $Z$-spin on all $N-1$ qubits of $m$. The coefficient $c(k_{N-2},j,s)$ is given in the following:

\begin{equation}
    c(k_{N-2},j,s) = \begin{cases}\sqrt{\frac{s}{2s+1}} & \text{ if } s = j + \frac{1}{2} \text{ and } j = k_{N-2} - \frac{1}{2} \\ -\sqrt{\frac{s+1}{2s+1}} &\text{ if } s = j - \frac{1}{2} \text{ and } j = k_{N-2} + \frac{1}{2} \\ 0 & \text{ otherwise }\end{cases}.
\end{equation}
In particular, noting that $j$ may be greater than or less than $k_{N-2}$ by $\frac{1}{2}$ and $s$ may be greater than or less than $j$ by $\frac{1}{2}$, we are seeing that $\bra{\psi^-}_{A_NA_{N+1}}\ket{k,j,s,m}$ is zero if $k_{N-2}$ differs from $s$ i.e. if $s = k_{N-2} \pm 1$. From this, one easily finds the coefficients of $\rho^{-1/2}\sigma^{(N)}\rho^{-1/2}$ in the Schur basis:

\begin{align}
    \bra{k',j',s',m'}\rho^{-1/2}\sigma^{(N)}\rho^{-1/2}\ket{k,j,s,m} &= \frac{(\lambda(j',s')\lambda(j,s))^{-1/2}}{2^{N-1}}c(k_{N-2}',j',s')c(k_{N-2},j,s)\delta_{kk'}\delta_{mm'}\label{firstLine} \\
    &= \frac{(\lambda(j',s')\lambda(j,s))^{-1/2}}{2^{N-1}}c(k_{N-2}',j',s')c(k_{N-2},j,s)\delta_{kk'}\delta_{mm'}\delta_{ss'}
\end{align}
where, going into the second line, we have simply added in $\delta_{ss'}$, which we are allowed to do because Equation \eqref{firstLine} shows that the expression vanishes unless $k_{N-2} = k_{N-2}'$ (from $\delta_{kk'}$) and $k_{N-2} = s$ (from $c(k_{N-2},j,s)$) and $k_{N-2}' = s'$ (from $c(k_{N-2}',j',s')$). Note also that in this expression we have assumed that neither $\ket{k,j,s,m}$ nor $\ket{k',j',s',m'}$ are maximally entangled states i.e. $s \neq \frac{N+1}{2}$ and $s' \neq \frac{N+1}{2}$ since then $(\lambda(j',s')\lambda(j,s))^{-1/2}$ would be undefined.

We therefore find that $\Pi_N\ket{k,j,s,m} = 0$ if $k_{N-2} \neq s$ and, if $k_{N-2} = s$, we find the action of $\Pi_N$ separately on $\ket{k,j,s,m}$ based on whether $j = s - 1/2$ or $j = s + 1/2$:

\begin{align}
    \Pi_N\ket{\tilde{k},s,s-\frac{1}{2},s,m} &= \frac{c(s,s-\frac{1}{2},s)^2}{2^{N-1}\lambda(s-\frac{1}{2},s)}\ket{\tilde{k},s,s-\frac{1}{2},s,m} + \frac{c(s,s-\frac{1}{2},s)c(s,s+\frac{1}{2},s)}{2^{N-1}\sqrt{\lambda(s-\frac{1}{2},s)\lambda(s+\frac{1}{2},s)}}\ket{\tilde{k},s,s+\frac{1}{2},s,m}\\
    \Pi_N\ket{\tilde{k},s,s+\frac{1}{2},s,m} &=\frac{c(s,s+\frac{1}{2},s)^2}{2^{N-1}\lambda(s+\frac{1}{2},s)}\ket{\tilde{k},s,s+\frac{1}{2},s,m} + \frac{c(s,s-\frac{1}{2},s)c(s,s+\frac{1}{2},s)}{2^{N-1}\sqrt{\lambda(s-\frac{1}{2},s)\lambda(s+\frac{1}{2},s)}}\ket{\tilde{k},s,s-\frac{1}{2},s,m}
\end{align}
recalling that $\tilde{k}$ denotes $k_1, ..., k_{N-3}$. We can see that the states $\ket{\tilde{k},s,s\pm\frac{1}{2},s,m}$ span a 2-dimensional invariant subspace under $\Pi_N$. In this subspace, $\Pi_N$ takes the form
\begin{equation}
    \begin{pmatrix} \alpha(s)^2 & \alpha(s)\beta(s) \\ \alpha(s)\beta(s) & \beta(s)^2\end{pmatrix} \text{ for } \alpha(s) = \frac{c(s,s-\frac{1}{2},s)}{\sqrt{2^{N-1}\lambda(s-\frac{1}{2},s)}} \text{ and } \beta(s) = \frac{c(s,s+\frac{1}{2},s)}{\sqrt{2^{N-1}\lambda(s+\frac{1}{2},s)}}.
\end{equation}
This is then an easily diagonalised matrix, having eigenvalues $0$ and $\alpha(s)^2 + \beta(s)^2$, and we conclude that $\Pi_N$ has the following eigenvectors and corresponding eigenvalues.
\begin{alignat}{2}
    &\ket{k,j,s,m} \text{ with } s = \frac{N+1}{2} \text{ i.e. maximal spin states} \hspace*{1cm} &&\text{ have eigenvalue } \frac{1}{N}\\
    &\ket{k,j,s,m} \text{ with } s < \frac{N+1}{2} \text{ and } k_{N-2} \neq s \hspace*{1cm} &&\text{ have eigenvalue } 0\\
    &\frac{\beta(s)}{\sqrt{\alpha(s)^2+\beta(s)^2}}\ket{\tilde{k},s,s-\frac{1}{2},s,m} - \frac{\alpha(s)}{\sqrt{\alpha(s)^2+\beta(s)^2}}\ket{\tilde{k},s,s+\frac{1}{2},s,m} \hspace*{1cm} &&\text{ have eigenvalue } 0\\
    &\frac{\alpha(s)}{\sqrt{\alpha(s)^2+\beta(s)^2}}\ket{\tilde{k},s,s-\frac{1}{2},s,m} + \frac{\beta(s)}{\sqrt{\alpha(s)^2+\beta(s)^2}}\ket{\tilde{k},s,s+\frac{1}{2},s,m} \hspace*{1cm} &&\text{ have eigenvalue } \alpha(s)^2 + \beta(s)^2.
\end{alignat}
Finally, because $\Pi_i = SWAP_{A_iA_N}\Pi_NSWAP_{A_iA_N}$, as mentioned, the eigenvectors for $\Pi_i$ can be found by acting on those of $\Pi_N$ with $SWAP_{A_iA_N}$, and they will have the same eigenvalues.

\subsection{Probabilistic Port-Based Teleportation}

\subsubsection{pPBT with Maximally Entangled Resource State}\label{pPBTMESPOVM}

We recall that in this case we deal with the POVM operators $
    \Pi_i = \ket{\psi^-}\bra{\psi^-}_{A_iA_{N+1}}\otimes\tilde{\Theta}_{i\bar{A_i}}$ for $i = 1, ..., N$ and $\Pi_{N+1} = I - \sum_{i=1}^N\Pi_i$,
where $\tilde{\Theta}_{i\bar{A}_i}$ denotes the operator $\tilde{\Theta}_i$ acting on the qubits $\bar{A}_i = A_1...A_{i-1}A_{i+1}...A_N$. This is an $N-1$ qubit operator defined as

\begin{equation}
    \tilde{\Theta}_i = \frac{1}{2^{N-1}}\sum_{s=s_{min}}^{(N-1)/2}\frac{1}{\lambda^+_{s+1/2}}\mathbb{I}(s)
\end{equation}
where $s_{min} = \begin{cases} 0 & \text{ if } N-1 \text{ is even} \\ \frac{1}{2} & \text{ if } N-1 \text{ is odd} \end{cases}$, $ \lambda^+_j = \frac{1}{2^N}\left(\frac{N}{2}+j+1\right) $
and $\mathbb{I}(s)$ is the projector onto the subspace of the $N-1$ qubits with total spin $s$.

Using the same notation as in the previous subsection, and using the same strategy of diagonalising $\Pi_N$ first and then using $\Pi_i = SWAP_{A_iA_N}\Pi_NSWAP_{A_iA_N}$, we find in this case that

\begin{equation}
    \bra{k',j',s',m'}\Pi_N\ket{k,j,s,m} = \frac{c(k_{N-2}',j',s')c(k_{N-2},j,s)}{2^{N-1}\lambda(s+\frac{1}{2},s)}\delta_{kk'}\delta_{mm'}\delta_{ss'}.
\end{equation}
We see that $\Pi_N$ is annihilating Schur states $\ket{k,j,s,m}$ for which $k_{N-2} \neq s$ i.e. if $k_{N-2} = s \pm 1$. We find the action of $\Pi_N$ on states with $k_{N-2} = s$ is as follows:

\begin{align}
    \Pi_N\ket{\tilde{k},s,s-\frac{1}{2},s,m} &= \frac{1}{2^{N-1}\lambda(s+\frac{1}{2},s)(2s+1)}\left[s\ket{\tilde{k},s,s-\frac{1}{2},s,m}-\sqrt{s(s+1)}\ket{\tilde{k},s,s+\frac{1}{2},s,m}\right]\\
    \Pi_N\ket{\tilde{k},s,s+\frac{1}{2},s,m} &= \frac{1}{2^{N-1}\lambda(s+\frac{1}{2},s)(2s+1)}\left[(s+1)\ket{\tilde{k},s,s+\frac{1}{2},s,m}-\sqrt{s(s+1)}\ket{\tilde{k},s,s-\frac{1}{2},s,m}\right]
\end{align}
and again the states $\ket{\tilde{k},s,s\pm\frac{1}{2},s,m}$ form a 2D invariant subspace under $\Pi_N$. This time, $\Pi_N$ takes the following form in this subspace.

\begin{equation}
    \frac{1}{2^{N-1}\lambda(s+\frac{1}{2},s)(2s+1)}\begin{pmatrix}s & -\sqrt{s(s+1)} \\ -\sqrt{s(s+1)} & s+1\end{pmatrix}
\end{equation}
which is again a matrix that is easily diagonalised. We conclude that $\Pi_N$ has the following eigenvectors and corresponding eigenvalues.

\begin{alignat}{2}
&\ket{k,j,s,m} \text{ with } k_{N-2} \neq s && \hspace*{1cm} \text{ have eigenvalue 0}\\
&\sqrt{\frac{s+1}{2s+1}}\ket{\tilde{k},s,s-\frac{1}{2},s,m} + \sqrt{\frac{s}{2s+1}}\ket{\tilde{k},s,s+\frac{1}{2},s,m} && \hspace*{1cm} \text{ have eigenvalue 0}\\
&\sqrt{\frac{s}{2s+1}}\ket{\tilde{k},s,s-\frac{1}{2},s,m} - \sqrt{\frac{s+1}{2s+1}}\ket{\tilde{k},s,s+\frac{1}{2},s,m} && \hspace*{1cm} \text{ have eigenvalue } \frac{1}{2^{N-1}\lambda(s+\frac{1}{2},s)}.
\end{alignat}
Again, the eigenvectors of $\Pi_i$ for $i = 1, ..., N-1$ may be found by applying $SWAP_{A_iA_N}$ to each of the eigenvectors of $\Pi_N$, giving the same corresponding eigenvalue, because we have $\Pi_i = SWAP_{A_iA_N}\Pi_NSWAP_{A_iA_N}$.

We must, however, make extra steps in this case because $\Pi_{N+1} = I - \sum_{i=1}^N\Pi_i$ requires diagonalisation also. We compute

\begin{align}
    \sum_{i=1}^N\Pi_i &= \sum_{s=s_{min}}^{\frac{N-1}{2}}\frac{1}{\lambda^+_{s+\frac{1}{2}}}\sum_{i=1}^N\frac{\ket{\psi^-}\bra{\psi^-}_{A_iA_{N+1}}\otimes \mathbb{I}(s)_{\bar{A}_i}}{2^{N-1}}\\
    &=\sum_{s=s_{min}}^{\frac{N-1}{2}}\frac{1}{\lambda^+_{s+\frac{1}{2}}}\rho(s)
\end{align}
where $\rho(s) = \mathbb{I}(s)_{AA_{N+1}}\rho\mathbb{I}(s)_{AA_{N+1}}$, and here $\mathbb{I}(s)_{AA_{N+1}}$ is the projector onto the subspace of the $N+1$ qubits with total spin $s$\footnote{This latter step may be seen by, for example, showing $\mathbb{I}(s)\left(\ket{\psi^-}\bra{\psi^-}_{A_NA_{N+1}}\otimes I_{\bar{A}_N}\right)\mathbb{I}(s) = \ket{\psi^-}\bra{\psi^-}_{A_NA_{N+1}}\otimes \mathbb{I}(s)_{\bar{A}_N}$ by showing equality of their matrix elements in the Schur basis, before showing equality of every term in the sum by acting with $SWAPs$.}. Knowing the eigenbasis for $\rho$ is $\ket{k,j,s,m}$, we may conclude that $\Pi_{N+1}$ has eigenvectors and corresponding eigenvalues as follows.

\begin{alignat}{2}
    &\ket{k,j,s,m} \text{ with } s = j + \frac{1}{2} \hspace*{2cm} &&\text{ have eigenvalue } 1 - \frac{\lambda(s-\frac{1}{2},s)}{\lambda(s+\frac{1}{2},s)}\\
    &\ket{k,j,s,m} \text{ with } s = j - \frac{1}{2} \hspace*{2cm} &&\text{ have eigenvalue } 0.
\end{alignat}

\subsubsection{pPBT with Optimised Resource State}\label{pPBTOptiPOVM}

Lastly, let us diagonalise the POVM operators used in the case of pPBT with optimised resource state, which we recall are $
    \Pi_i = (O^{-1}_A\otimes I_{A_{N+1}})\left(\ket{\psi^-}\bra{\psi^-}_{A_iA_{N+1}}\otimes \tilde{\Theta}_{i\bar{A}_i}\right)(O^{-1}_A\otimes I_{A_{N+1}})\; \text{ for } \;i = 1, ..., N \;\text{ and }\; \Pi_{N+1} = I - \sum_{i=1}^N\Pi_i$
where $A = A_1...A_N$ denotes Alice's ports and
\begin{align}
    O &= \sum_{j = j_{min}}^{N/2}\sqrt{\nu(j)}\mathbb{I}(j)_A \;\;\;\text{ for }\;\;\; \nu(j) = \frac{2^Nh(N)(2j+1)}{g^{[N]}(j)} \;\;\;\text{ and }\;\;\; j_{min} = \begin{cases}0 & \text{ if } N \text{ is even} \\ \frac{1}{2} & \text{ if } N \text{ is odd}\end{cases}\\
    \tilde{\Theta}_i &= \sum_{s=s_{min}}^{(N-1)/2}u(s)\mathbb{I}(s)\;\;\;\;\;\;\;\text{ for }\;\;\;u(s) = \frac{2^{N+1}h(N)(2s+1)}{Ng^{[N-1]}(s)}
\end{align}
where $\mathbb{I}(j)$ is the projector onto the subspace of $N$ qubits with spin $j$, $h(N) = \frac{6}{(N+1)(N+2)(N+3)}$ and $g^{[N]}(j) = \frac{(2j+1)N!}{(\frac{N}{2}-j)!(\frac{N}{2}+1+j)!}$. Using the strategy once again of diagonalising $\Pi_N$ first before employing $\Pi_i = SWAP_{A_iA_N}\Pi_NSWAP_{A_iA_N}$, we find this time that

\begin{equation}
    \bra{k',j',s',m'}\Pi_N\ket{k,j,s,m} = (\nu(j')\nu(j))^{-1/2}\:u(s)\:c(k'_{N-2},j',s')\:c(k_{N-2},j,s)\:\delta_{kk'}\delta_{mm'}\delta_{ss'},
\end{equation}
finding again that any $\ket{k,j,s,m}$ is annihilated with $k_{N-2} \neq s$. The action of $\Pi_N$ on states $\ket{k,j,s,m}$ with $k_{N-2} = s$ is then

\begin{align}
    \Pi_N\ket{\tilde{k},s,s-\frac{1}{2},s,m} &= u(s)\left[\frac{c(s,s-\frac{1}{2},s)^2}{\nu(s-\frac{1}{2})}\ket{\tilde{k},s,s-\frac{1}{2},s,m}+\frac{c(s,s-\frac{1}{2},s)c(s,s+\frac{1}{2},s)}{\sqrt{\nu(s+\frac{1}{2})\nu(s-\frac{1}{2})}}\ket{\tilde{k},s,s+\frac{1}{2},s,m}\right]\\
    \Pi_N\ket{\tilde{k},s,s+\frac{1}{2},s,m} &= u(s)\left[\frac{c(s,s+\frac{1}{2},s)^2}{\nu(s+\frac{1}{2})}\ket{\tilde{k},s,s+\frac{1}{2},s,m}+\frac{c(s,s-\frac{1}{2},s)c(s,s+\frac{1}{2},s)}{\sqrt{\nu(s+\frac{1}{2})\nu(s-\frac{1}{2})}}\ket{\tilde{k},s,s-\frac{1}{2},s,m}\right].
\end{align}
In this 2D invariant subspace, $\Pi_N$ takes the form

\begin{equation}
    u(s)\begin{pmatrix}
       \gamma(s)^2 & \gamma(s)\delta(s) \\ \gamma(s)\delta(s) & \delta(s)^2 
    \end{pmatrix} \;\;\text{ where }\;\;
\gamma(s) = \frac{c(s,s-\frac{1}{2},s)}{\sqrt{\nu(s-\frac{1}{2})}} \;\;\text{ and }\;\; \delta(s) = \frac{c(s,s+\frac{1}{2},s)}{\sqrt{\nu(s+\frac{1}{2})}}.
\end{equation}
We thus conclude that $\Pi_N$ has eigenvectors with corresponding eigenvalues
\begin{alignat}{2}
&\ket{k,j,s,m} \text{ with } k_{N-2} \neq s \hspace*{0.8cm}&&\text{ have eigenvalue } 0\\
&\frac{\delta(s)}{\sqrt{\gamma(s)^2+\delta(s)^2}}\ket{\tilde{k},s,s-\frac{1}{2},s,m} - \frac{\gamma(s)}{\sqrt{\gamma(s)^2+\delta(s)^2}}\ket{\tilde{k},s,s+\frac{1}{2},s,m} \hspace*{0.8cm}&&\text{ have eigenvalue } 0\\
&\frac{\gamma(s)}{\sqrt{\gamma(s)^2+\delta(s)^2}}\ket{\tilde{k},s,s-\frac{1}{2},s,m} + \frac{\delta(s)}{\sqrt{\gamma(s)^2+\delta(s)^2}}\ket{\tilde{k},s,s+\frac{1}{2},s,m} \hspace*{0.8cm}&&\text{ have eigenvalue } u(s)\left(\gamma(s)^2+\delta(s)^2\right).
\end{alignat}
Once again, the eigenvectors and corresponding eigenvalues of $\Pi_i$ are found by acting with $SWAP_{A_iA_N}$ on the eigenvectors of $\Pi_N$.

Finally, we must diagonalise $\Pi_{N+1} = I - \sum_{i=1}^N\Pi_i$. We compute

\begin{align}
    \sum_{i=1}^N\Pi_i &= \left(O_A^{-1}\otimes I_{A_{N+1}}\right)\sum_{i=1}^N\left(\ket{\psi^-}\ket{\psi^-}_{A_iA_{N+1}}\otimes\tilde{\Theta}_{i\bar{A}_i}\right)\left(O_A^{-1}\otimes I_{A_{N+1}}\right)\\
    &=2^{N-1}\left(O_A^{-1}\otimes I_{A_{N+1}}\right)\sum_{s=s_{min}}^{\frac{N-1}{2}}u(s)\rho(s)\left(O_A^{-1}\otimes I_{A_{N+1}}\right)
\end{align}
This manifestly has an eigenbasis $\ket{k,j,s,m}$ and so we find that $\Pi_{N+1}$ has the set of eigenvectors $\ket{k,j,s,m}$ with eigenvalues

\begin{equation}
    1 - \frac{2^{N-1}\lambda(j,s)u(s)}{\nu(j)}.
\end{equation}

\section{Algorithms for Implementing Port-Based Teleportation}\label{algos}

We now present algorithms for the POVMs of dPBT and pPBT. Again, only three cases require consideration to cover the four regimes because the same POVM may be used for both maximally entangled resource state and optimised resource state in the case of dPBT \cite{leditzky2022optimality}. As such, deterministic port-based teleportation is handled in one go in Section \ref{dPBTAlgo}, pPBT with maximally entangled resource state is handled in Section \ref{pPBTMESAlgo}, and pPBT with optimised resource state is handled in Section \ref{pPBTOptAlgo}. Additionally, we provide an alternative protocol for pPBT with maximally entangled resource state in Section \ref{alternativepPBT}. This algorithm is optimised for practicality for near-term devices, where we use no amplitude amplification at all. This is at the expense of success probability - this algorithm unfortunately has an asymptotic success probability of $\frac{1}{4}$, but in many cases this is acceptable, for example if one has some (small) constant number of the input state. This section is independent of the previous section - all results proved in the previous section that are required here will be quoted. We do, however, suggest that the following sections are read sequentially, since there are common themes between each algorithm.

We will use in every instance the ($N+1$)-qubit inverse coupling stage of the Schur transform (which recall is what many people simply refer to as the Schur transform)\footnote{We go into the different definitions of the Schur transform, and provide useful translations between them in Appendix \ref{preMapping}, which will likely be of use to those who come from different backgrounds.}, where Schur states $\ket{k,j,s,m}$ are mapped to encodings of their various spins $\ket{k}\ket{j}\ket{s}\ket{m}$. In each case, for this operation, one could either choose to use the algorithm of \cite{wills2023generalised}, which is slower but uses fewer ancillas, or \cite{bacon2005quantum}, which is faster but uses more ancillas. We mention one largely unimportant difference, which is that in the former case the register $\ket{k}$ is an encoding containing $n-2$ qubits, and in the latter case it contains $\mathcal{O}(n\log(n))$ qubits.

We will not mention the differences between using these two algorithms for this operation for the remainder of this section to avoid confusion. We will simply say that we will perform the (inverse coupling stage for the) Schur transform. We will delegate the analysis of the various possibilities to Appendix \ref{complexityAnalysis}, where explicit gate complexities and ancilla counts will be discussed, and full tables showing these values may be found in the introduction.

We will repeatedly use the technique of oblivious amplitude amplification as described in Theorem 28 of \cite{gilyen2019quantum}. We state this now as in that work.

\begin{theorem}[Theorem 28 of \cite{gilyen2019quantum} on Robust Oblivious Amplitude Amplification]\label{AAThm}
    Let $n \in \mathbb{N}_+$ be odd, let $\epsilon \in \mathbb{R}_+$, let $U$ be a unitary, let $\tilde{\Pi}$, $\Pi$ be orthogonal projectors, and let $W: \text{img}(\Pi) \to \text{img}(\tilde{\Pi})$ be an isometry, such that
    \begin{equation}
        \left|\left|\sin\left(\frac{\pi}{2n}\right)W\ket{\Psi} - \tilde{\Pi}U\ket{\Psi}\right|\right| \leq \epsilon
    \end{equation}
    for all $\ket{\Psi} \in \text{img}(\Pi)$. Then we can construct a unitary $\tilde{U}$ such that for all $\ket{\Psi} \in \text{img}(\Pi)$,
    \begin{equation}
        \left|\left|W\ket{\Psi} - \tilde{\Pi}\tilde{U}\ket{\Psi}\right|\right| \leq 2n\epsilon
    \end{equation}
    which uses a single ancilla qubit, with $n$ uses of $U$ and $U^\dagger$, $n$ uses of $C_\Pi NOT$ and $n$ uses of $C_{\tilde{\Pi}} NOT$ gates and $n$ single qubit gates, where
    \begin{equation}
        C_\Pi NOT = X \otimes \Pi + I \otimes (I-\Pi).
    \end{equation}
\end{theorem}
We note that in the case of $\epsilon = 0$, $W$ has the same action as $\tilde{\Pi}\tilde{U}$ on $\text{img}(\Pi)$, and so in particular is an isometry mapping into $\text{img}(\tilde{\Pi})$. As a result, in the case $\epsilon = 0$, $\tilde{U}$ must map into $\text{img}(\tilde{\Pi})$, since otherwise the non-trivial action of $\tilde{\Pi}$ makes $\tilde{\Pi}\tilde{U}$ not an isometry. For the purposes of explaining the algorithms in this section, we will assume everywhere that $\epsilon = 0$, i.e. all of our gates can be performed perfectly, and will consider inaccuracies in the complexity analysis of Appendix \ref{complexityAnalysis}.

We also note that amplitude amplification, while an extraordinarily useful and flexible tool for algorithm design, is quite demanding in practical terms, and so one of our aims will be to keep the number of rounds of amplitude amplification to a minimum i.e. keeping the integer $n$ low.

With a description of oblivious amplitude amplification given, we can state that the algorithms in each case share the same broad framework. For the square-root measurement of dPBT, as implemented in Section \ref{dPBTAlgo}, the algorithms for pPBT with maximally entangled resource state in Sections \ref{pPBTMESAlgo} and \ref{alternativepPBT} and the algorithm for pPBT with optimised resource state in Section \ref{pPBTOptAlgo}, the steps of each may be loosely summarised as follows.

\begin{enumerate}
    \item Rotate to a shared eigenbasis of the POVM elements.
    \item Using an ancillary qubit $\ket{r}$, attach the eigenvalues of the POVM elements in the subspace $\ket{r=0}$, rotating everything we don't want into some ``junk'' subspace $\ket{r=1}$
    \item Amplify the $\ket{r=0}$ subspace with oblivious amplitude amplification.
\end{enumerate}

The exact meaning of each of these steps will become clearer in the coming sections, but we list them here with the intention of commenting on their potential for generalisation. In certain cases, the method for implementing the square-root measurement via the Petz recovery \cite{gilyen2022quantum} runs inefficiently, including dPBT, due to the exponentially small positive eigenvalues on the ensemble average state. To implement such an operation efficiently, one is obliged to exploit the shared symmetries of the state ensemble in question, as we do here. These symmetries make both the rotation of Step 1, above, efficiently implementable, and also make the eigenvalues highly degenerate, making the second step feasible. Our approach can therefore be seen as a potentially fruitful framework for implementing the square-root measurement for other highly symmetric state ensembles exponentially more efficiently than the Petz recovery approach.

We lastly comment that we will use the same notation as the previous section, where we denote $(N+1)$-qubit Schur states by $\ket{k_1, k_2, ..., k_{N-2},j,s,m}$, where $k_1$ is the spin on the first two qubits, $k_2$ is the spin on the first three qubits, and so on, until $k_{N-2}$ is the spin on the first $N-1$ qubits (which for us is $A_1...A_{N-1}$) and $j$ is the spin on the first $N$ qubits (which for us is all of Alice's ports $A = A_1...A_N$). $s$ is then the spin on all $N+1$ qubits (Alice's ports and the input state, $AA_{N+1}$) and $m$ is the $Z$-spin on all $N+1$ qubits. Again, when this notation is too cumbersome, we will abbreviate it to $\ket{k,j,s,m}$ where $k$ stands for $k_1, ..., k_{N-2}$, and in cases where we want to address $k_{N-2}$ but none of the other $k$'s, we will write $\ket{\tilde{k},k_{N-2},j,s,m}$ where $\tilde{k}$ stands for $k_1, ..., k_{N-3}$. We emphasise the fact that each consecutive spin may only differ from each other by $1/2$ and may not fall below zero as a consequence of the rules of addition of angular momentum in Equation \eqref{AMRules}. In particular, to have $s = \frac{N+1}{2}$, we must have $k_1 = 1$, $k_2 = \frac{3}{2}$, $k_3 = 2$, and so on, up to $j = \frac{N}{2}$.

\subsection{Deterministic Port-Based Teleportation}\label{dPBTAlgo}

Starting with a state $\ket{\psi}$ on $N+1$ qubits, we attach an ancillary register $\ket{i}$ of dimension $N$ (which, in particular, may be made up of $\lceil \log(N)\rceil$ qubits) with an orthonormal basis $\{\ket{i}\}_{i=1}^N$. The goal is to implement the operation

\begin{equation}
    \ket{\psi}\ket{1}_i \mapsto \sum_{i=1}^N\sqrt{\Pi_i}\ket{\psi}\ket{i}
\end{equation}
where we recall here the eigendecomposition of each $\Pi_i$. We have $\Pi_i = SWAP_{A_iA_N}\Pi_NSWAP_{A_iA_N}$, and $\Pi_N$ has the set of eigenvectors with corresponding eigenvalues

\begin{alignat}{2}
    &\ket{k,j,s,m} \text{ with } s = \frac{N+1}{2} \text{ i.e. maximal spin states} \hspace*{1cm} &&\text{ have eigenvalue } \frac{1}{N}\\
    &\ket{k,j,s,m} \text{ with } s < \frac{N+1}{2} \text{ and } k_{N-2} \neq s \hspace*{1cm} &&\text{ have eigenvalue } 0\\
    &\frac{\beta(s)}{\sqrt{\alpha(s)^2+\beta(s)^2}}\ket{\tilde{k},s,s-\frac{1}{2},s,m} - \frac{\alpha(s)}{\sqrt{\alpha(s)^2+\beta(s)^2}}\ket{\tilde{k},s,s+\frac{1}{2},s,m} \hspace*{1cm} &&\text{ have eigenvalue } 0\\
    &\frac{\alpha(s)}{\sqrt{\alpha(s)^2+\beta(s)^2}}\ket{\tilde{k},s,s-\frac{1}{2},s,m} + \frac{\beta(s)}{\sqrt{\alpha(s)^2+\beta(s)^2}}\ket{\tilde{k},s,s+\frac{1}{2},s,m} \hspace*{1cm} &&\text{ have eigenvalue } \alpha(s)^2 + \beta(s)^2.
\end{alignat}
where we have
\begin{equation}
    \alpha(s) = \sqrt{\frac{2s}{(2s+1)(N+1-2s)}}, \hspace*{1cm} \beta(s) = -\sqrt{\frac{4s+4}{(2s+1)(N+3+2s)}}
\end{equation}
from which we compute
\begin{equation}
    \alpha(s)^2 + \beta(s)^2 = \frac{4(N+1)}{(N+1-2s)(N+3+2s)}.
\end{equation}
The first step of our algorithm will be to create the superposition
\begin{equation}
    \ket{\psi}\ket{1}_i \mapsto \frac{1}{\sqrt{N}}\sum_{i=1}^N\ket{\psi}\ket{i}
\end{equation}
after which we apply the operation
\begin{equation}
    C-SWAP = \sum_{i=1}^N SWAP_{A_iA_N}\otimes \ket{i}\bra{i}
\end{equation}
to obtain
\begin{equation}
\frac{1}{\sqrt{N}}\sum_{i=1}^N\ket{\psi_i}\ket{i}
\end{equation}
where $\ket{\psi_i} = SWAP_{A_iA_N}\ket{\psi}$. Then, the idea is to implement $\sqrt{\Pi_N}$ on the $\ket{\psi_i}$, before applying $C-SWAP$ again. In order to implement $\sqrt{\Pi_N}$, we will need one further ancillary qubit in the register $\ket{r}$, which we initialise to $\ket{0}$\footnote{This register is what allows us to create a block-encoding of $\sqrt{\Pi_N}$. Note one advantage of our algorithm is that it uses only one ancillary qubit for the block-encoding as opposed to the logarithmic number used in \cite{fei2023efficient}.}.

Now that we have $\frac{1}{\sqrt{N}}\sum_{i=1}^N\ket{\psi_i}\ket{i}\ket{0}_r$, the idea is simple. Rotate to the eigenbasis of $\Pi_N$, put the eigenvalues of $\sqrt{\Pi_N}$ onto the corresponding eigenvectors, rotating everything we don't want into $\ket{r=1}$, and lastly use amplitude amplification to amplify the $\ket{r=0}$ space.

To do this explicitly, we need a little more notation so that we can capture more of the eigenvectors simultaneously. With $s < \frac{N+1}{2}$, we make the following definitions:

\begin{align}
    \ket{\tilde{k},-,-,s,m} &= \ket{\tilde{k},s-1,s-\frac{1}{2},s,m}\label{rotEqn80}\\
    \ket{\tilde{k},+,-,s,m} &= \ket{\tilde{k},s+1,s+\frac{1}{2},s,m}\\
    \ket{\tilde{k},-,+,s,m} &= \frac{\beta(s)}{\sqrt{\alpha(s)^2+\beta(s)^2}}\ket{\tilde{k},s,s-\frac{1}{2},s,m} - \frac{\alpha(s)}{\sqrt{\alpha(s)^2+\beta(s)^2}}\ket{\tilde{k},s,s+\frac{1}{2},s,m}\label{rotEqn1}\\
    \ket{\tilde{k},+,+,s,m} &= \frac{\alpha(s)}{\sqrt{\alpha(s)^2 + \beta(s)^2}}\ket{\tilde{k},s,s-\frac{1}{2},s,m} + \frac{\beta(s)}{\sqrt{\alpha(s)^2 + \beta(s)^2}}\ket{\tilde{k},s,s+\frac{1}{2},s,m}\label{rotEqn2}
\end{align}
which recall have respective eigenvalues $0,0,0$ and $\alpha(s)^2 + \beta(s)^2$ under $\Pi_N$. With $s = \frac{N+1}{2}$ we simply define
\begin{equation}
    \ket{\tilde{k},+,+,s,m} = \ket{\tilde{k},\frac{N-1}{2},\frac{N}{2},\frac{N+1}{2},m}
\end{equation}
which have eigenvalues $\frac{1}{N}$ under $\Pi_N$. Recall these are the maximal spin states and may only differ by their value of $m$. Note that in this latter expression, the choice of $(+,+)$ is completely arbitrary, we just make some notational choice so that $\ket{\psi_i}$ may be completely written out in terms of the eigenbasis of $\Pi_N$:
\begin{equation}
    \ket{\psi_i} = \sum_{\tilde{k},(a,b)\in \{\pm\}^2,s,m}c^{(i)}_{\tilde{k},a,b,s,m}\ket{\tilde{k},a,b,s,m}
\end{equation}
where we are only summing over the indices $(\tilde{k},a,b,s,m)$ that are allowed i.e. that produce valid eigenstates. $a$ and $b$ are symbols each being summed over the set $\{\pm\}$. 

We can see that it won't be too hard to rotate to the eigenbasis of $\Pi_N$. Indeed, we need only un-compute the Schur coupling, and then do one simple rotation to straighten out the coefficients in Equations \eqref{rotEqn1} and \eqref{rotEqn2}. Indeed, we start by doing the inverse of the coupling stage of the Schur transform, which again is what many people simply refer to as the Schur transform, thus mapping each eigenvector in the following way:
\begin{align}
    \ket{\tilde{k},-,-,s,m}&\mapsto \ket{\tilde{k}}\ket{s-1}\ket{s-\frac{1}{2}}\ket{s}\ket{m}\label{rotEqn86}\\
    \ket{\tilde{k},+,-,s,m}&\mapsto \ket{\tilde{k}}\ket{s+1}\ket{s+\frac{1}{2}}\ket{s}\ket{m}\\
    \ket{\tilde{k},-,+,s,m}&\mapsto\frac{\beta(s)}{\sqrt{\alpha(s)^2+\beta(s)^2}}\ket{\tilde{k}}\ket{s}\ket{s-\frac{1}{2}}\ket{s}\ket{m} - \frac{\alpha(s)}{\sqrt{\alpha(s)^2+\beta(s)^2}}\ket{\tilde{k}}\ket{s}\ket{s+\frac{1}{2}}\ket{s}\ket{m}\label{invSchurEqn1}\\\label{invSchurEqn2}
    \ket{\tilde{k},+,+,s,m}&\mapsto\frac{\alpha(s)}{\sqrt{\alpha(s)^2 + \beta(s)^2}}\ket{\tilde{k}}\ket{s}\ket{s-\frac{1}{2}}\ket{s}\ket{m} + \frac{\beta(s)}{\sqrt{\alpha(s)^2 + \beta(s)^2}}\ket{\tilde{k}}\ket{s}\ket{s+\frac{1}{2}}\ket{s}\ket{m}
\end{align}
if $s < \frac{N+1}{2}$, whereas if $s = \frac{N+1}{2}$, we just get
\begin{equation}
    \ket{\tilde{k},+,+,s,m}\mapsto\ket{\tilde{k}}\ket{\frac{N-1}{2}}\ket{\frac{N}{2}}\ket{\frac{N+1}{2}}\ket{m}.
\end{equation}
We may then perform a rotation acting on the second, third and fourth registers of these expressions, with the aim of just straightening out the coefficients in Equations \eqref{invSchurEqn1} and \eqref{invSchurEqn2}. Explicitly,
\begin{align}
    \ket{\frac{N-1}{2}}\ket{\frac{N}{2}}\ket{\frac{N+1}{2}} &\mapsto \ket{\frac{N-1}{2}}\ket{\frac{N}{2}}\ket{\frac{N+1}{2}}\label{detRot1}\\
    \ket{s-1}\ket{s-\frac{1}{2}}\ket{s} &\mapsto \ket{s-1}\ket{s-\frac{1}{2}}\ket{s}\\
    \ket{s+1}\ket{s+\frac{1}{2}}\ket{s} &\mapsto \ket{s+1}\ket{s+\frac{1}{2}}\ket{s}\\
    \ket{s}\ket{s-\frac{1}{2}}\ket{s} &\mapsto \frac{\beta(s)}{\sqrt{\alpha(s)^2+\beta(s)^2}}\ket{s}\ket{s-\frac{1}{2}}\ket{s} + \frac{\alpha(s)}{\sqrt{\alpha(s)^2+\beta(s)^2}}\ket{s}\ket{s+\frac{1}{2}}\ket{s}\\
    \ket{s}\ket{s+\frac{1}{2}}\ket{s} &\mapsto \frac{-\alpha(s)}{\sqrt{\alpha(s)^2+\beta(s)^2}}\ket{s}\ket{s-\frac{1}{2}}\ket{s} + \frac{\beta(s)}{\sqrt{\alpha(s)^2+\beta(s)^2}}\ket{s}\ket{s+\frac{1}{2}}\ket{s}\label{detRot5}
\end{align}
so that in total we have acted on the eigenstates of $\Pi_N$ as

\begin{align}
    \ket{\tilde{k},-,-,s,m}&\mapsto \ket{\tilde{k}}\ket{s-1}\ket{s-\frac{1}{2}}\ket{s}\ket{m}\label{firstMap}\\
    \ket{\tilde{k},+,-,s,m}&\mapsto \ket{\tilde{k}}\ket{s+1}\ket{s+\frac{1}{2}}\ket{s}\ket{m}\\
    \ket{\tilde{k},-,+,s,m}&\mapsto\ket{\tilde{k}}\ket{s}\ket{s-\frac{1}{2}}\ket{s}\ket{m}\\
    \ket{\tilde{k},+,+,s,m}&\mapsto\ket{\tilde{k}}\ket{s}\ket{s+\frac{1}{2}}\ket{s}\ket{m}.
\end{align}
for $s < \frac{N+1}{2}$, whereas if $s = \frac{N+1}{2}$, 
\begin{equation}
    \ket{\tilde{k},+,+,s,m}\mapsto\ket{\tilde{k}}\ket{\frac{N-1}{2}}\ket{\frac{N}{2}}\ket{\frac{N+1}{2}}\ket{m}\label{fifthMap}.
\end{equation}
Recalling that adjacent spins may only differ from each other by $1/2$, we note that given a fixed $s$, the value of the third register in Equations \eqref{firstMap} - \eqref{fifthMap} may only be $s \pm \frac{1}{2}$. Similarly, give some value $j$ for the third register, the second register may only take two values again: $j \pm \frac{1}{2}$. Since the encoding of one of two values may be done by a single qubit, we can therefore compress the second and third registers into one qubit each, mapping
\begin{align}
    \ket{s-1}\ket{s-\frac{1}{2}}\ket{s} &\mapsto \ket{-}\ket{-}\ket{s}\label{compressionDet1}\\
    \ket{s+1}\ket{s+\frac{1}{2}}\ket{s} &\mapsto \ket{+}\ket{-}\ket{s}\\
    \ket{s}\ket{s-\frac{1}{2}}\ket{s} &\mapsto \ket{-}\ket{+}\ket{s}\\
    \ket{s}\ket{s+\frac{1}{2}}\ket{s} &\mapsto \ket{+}\ket{+}\ket{s}
\end{align}
for $s < \frac{N+1}{2}$ and if $s = \frac{N+1}{2}$,
\begin{equation}
    \ket{\frac{N-1}{2}}\ket{\frac{N}{2}}\ket{\frac{N+1}{2}}\mapsto\ket{+}\ket{+}\ket{\frac{N+1}{2}}.\label{compressionDet2}
\end{equation}
In total, we have mapped
\begin{align}
    \ket{\psi_i} = \sum_{\tilde{k},a,b,s,m}c^{(i)}_{\tilde{k},a,b,s,m}\ket{\tilde{k},a,b,s,m} \mapsto \sum_{\tilde{k},a,b,s,m}c^{(i)}_{\tilde{k},a,b,s,m}\ket{\tilde{k}}\ket{a}\ket{b}\ket{s}\ket{m}.\label{eigenvectorMap}
\end{align}
It is now time to attach the eigenvalues of $\sqrt{\Pi_N}$. We write the block-encoding ancilla $\ket{r=0}$ back in, which of course has been there the entire time, and perform the following. If $s < \frac{N+1}{2}$, do
\begin{align}
    \ket{-}\ket{-}\ket{s}\ket{0}_r &\mapsto \ket{-}\ket{-}\ket{s}\ket{1}_r\label{eigenvalueAttachDet1}\\
    \ket{+}\ket{-}\ket{s}\ket{0}_r &\mapsto \ket{+}\ket{-}\ket{s}\ket{1}_r\\
    \ket{-}\ket{+}\ket{s}\ket{0}_r &\mapsto \ket{-}\ket{+}\ket{s}\ket{1}_r\\
    \ket{+}\ket{+}\ket{s}\ket{0}_r &\mapsto \sqrt{\alpha(s)^2+\beta(s)^2}\ket{+}\ket{+}\ket{s}\ket{0}_r + \sqrt{1-(\alpha(s)^2 + \beta(s)^2)}\ket{+}\ket{+}\ket{s}\ket{1}_r\label{nonTrivRot}
\end{align}
where we can check that $\alpha(s)^2+\beta(s)^2 \leq 1$ for $s < \frac{N+1}{2}$, actually achieving equality with $s = \frac{N-1}{2}$, and lastly for $s = \frac{N+1}{2}$,
\begin{equation}
    \ket{+}\ket{+}\ket{\frac{N+1}{2}}\ket{0}_r \mapsto \frac{1}{\sqrt{N}}\ket{+}\ket{+}\ket{\frac{N+1}{2}}\ket{0}_r + \sqrt{1-\frac{1}{N}}\ket{+}\ket{+}\ket{\frac{N+1}{2}}\ket{1}_r\label{maxSpinRot}
\end{equation}
where we see we have attached the exact eigenvalues we want for $\sqrt{\Pi_N}$. Therefore, now undoing the whole rotation to the eigenbasis of Equation \eqref{eigenvectorMap}, we have in total achieved
\begin{equation}
    \ket{\psi}\ket{1}_i\ket{0}_r \mapsto \frac{1}{\sqrt{N}}\sum_{i=1}^N\ket{\psi_i}\ket{i}\ket{0}_r \mapsto \frac{1}{\sqrt{N}}\sum_{i=1}^N\sqrt{\Pi_N}\ket{\psi_i}\ket{i}\ket{0}_r + \ket{*}\ket{1}_r
\end{equation}
where $\ket{*}$ is an unimportant state to us. Lastly, acting with $C-SWAP$ again gives
\begin{equation}
    \frac{1}{\sqrt{N}}\sum_{i=1}^N\sqrt{\Pi_i}\ket{\psi}\ket{i}\ket{0}_r + \ket{*}\ket{1}_r.
\end{equation}
It only remains to perform amplitude amplification to amplify the $\ket{r=0}$ eigenspace. Indeed, using the notation of Theorem \ref{AAThm}, which in turn uses the notation of \cite{gilyen2019quantum}, we set $\Pi = \ket{i=1}\bra{i=1}\otimes\ket{r=0}\bra{r=0}$ and $\tilde{\Pi} = \ket{r=0}\bra{r=0}$, and define the isometry $W: \text{img}(\Pi) \to \text{img}(\tilde{\Pi})$ by
\begin{equation}
    \ket{\psi}\ket{i=1}\ket{r=0} \mapsto \sum_{i=1}^N\sqrt{\Pi_i}\ket{\psi}\ket{i}\ket{r=0}.
\end{equation}
Up until now, we have constructed a unitary $U$ for which
\begin{equation}
    \tilde{\Pi}U\left(\ket{\psi}\ket{i=1}\ket{r=0}\right) = \frac{1}{\sqrt{N}}W\left(\ket{\psi}\ket{i=1}\ket{r=0}\right).
\end{equation}
Note that in an ideal world we would have $\frac{1}{\sqrt{N}} = \sin\left(\frac{\pi}{2n}\right)$ with $n$ an odd integer to apply Theorem \ref{AAThm}, and we usually will not. This can be dealt with in a straightforward way by just letting the number $c^*$ be the smallest number greater than or equal to 1 such that there is an odd integer $n$ for which $\frac{1}{c^*\sqrt{N}} = \sin\left(\frac{\pi}{2n}\right)$, and then modifying the rotations of Equations \eqref{nonTrivRot} and \eqref{maxSpinRot} so that we actually end up with
\begin{equation}
    \frac{1}{c^*\sqrt{N}}\sum_{i=1}^N\sqrt{\Pi_i}\ket{\psi}\ket{i}\ket{0}_r + \ket{*}\ket{1}_r.
\end{equation}
With this detail, we simply note that we need to perform $n = \mathcal{O}(\sqrt{N})$ rounds of amplitude amplification to produce the desired final state $\sum_{i=1}^N\sqrt{\Pi_i}\ket{\psi}\ket{i}\ket{r=0}$, which we can do via the prescription of Theorem 28 of \cite{gilyen2019quantum}, noting that in our case the choices of $\Pi$ and $\tilde{\Pi}$ make $C_\Pi NOT$ and $C_{\tilde{\Pi}} NOT$ easy to perform.

\subsection{Probabilistic Port-Based Teleportation}\label{pPBTAlgos}

\subsubsection{pPBT with Maximally Entangled Resource State - Algorithm 1}\label{pPBTMESAlgo}

In this case, we have a state on $N+1$ qubits and, because we have an $(N+1)$-element POVM, we attach an ancillary register $\ket{i}$ of dimension $N+1$, and the goal is to implement

\begin{equation}
    \ket{\psi}\ket{1}_i \mapsto \sum_{i=1}^{N+1}\sqrt{\Pi_i}\ket{\psi}\ket{i}.
\end{equation}
Again, we will use one ancillary qubit in the register $\ket{r}$ to do the block-encoding, initialised in the state $\ket{0}$, which we do not write for now. We will find that, because of the form of the POVM elements in this case, we are able to do this particularly efficiently - requiring only a (small) constant number of rounds of amplitude amplification, in contrast to the previous subsection. We recall that in this case, $\Pi_N$ has the eigendecomposition

\begin{alignat}{2}
&\ket{k,j,s,m} \text{ with } k_{N-2} \neq s && \hspace*{1cm} \text{ have eigenvalue 0}\\
&\sqrt{\frac{s+1}{2s+1}}\ket{\tilde{k},s,s-\frac{1}{2},s,m} + \sqrt{\frac{s}{2s+1}}\ket{\tilde{k},s,s+\frac{1}{2},s,m} && \hspace*{1cm} \text{ have eigenvalue 0}\\
&\sqrt{\frac{s}{2s+1}}\ket{\tilde{k},s,s-\frac{1}{2},s,m} - \sqrt{\frac{s+1}{2s+1}}\ket{\tilde{k},s,s+\frac{1}{2},s,m} && \hspace*{1cm} \text{ have eigenvalue } \sigma(s)
\end{alignat}
where $\sigma(s) = \frac{4}{N+3+2s}$ and for $i = 1, ..., N-1$, $\Pi_i = SWAP_{A_iA_N}\Pi_NSWAP_{A_iA_N}$. We then also have the eigendecomposition of $\Pi_{N+1}$, which is

\begin{alignat}{2}
    &\ket{k,j,s,m} \text{ with } s = j + \frac{1}{2} \hspace*{2cm} &&\text{ have eigenvalue } \tau(s)\\
    &\ket{k,j,s,m} \text{ with } s = j - \frac{1}{2} \hspace*{2cm} &&\text{ have eigenvalue } 0
\end{alignat}
where $\tau(s) = \frac{2(2s+1)}{N+3+2s}$.

In this case, we start by creating the superposition
\begin{equation}
    \ket{\psi}\ket{1}_i \mapsto \frac{1}{\sqrt{2N}}\sum_{i=1}^N\ket{\psi}\ket{i} + \frac{1}{\sqrt{2}}\ket{\psi}\ket{N+1}.
\end{equation}
This form of the initial superposition is chosen carefully so that we only need to do a constant number of rounds of amplitude amplification later on, making our algorithm quite efficient. We use a similar strategy to last time, now applying
\begin{equation}
    C-SWAP = \sum_{i=1}^NSWAP_{A_iA_N}\otimes\ket{i}\bra{i} + I\otimes \ket{N+1}\bra{N+1} \label{CSWAP}
\end{equation}
to get
\begin{equation}
    \frac{1}{\sqrt{2N}}\sum_{i=1}^N\ket{\psi_i}\ket{i} + \frac{1}{\sqrt{2}}\ket{\psi}\ket{N+1},\label{superposition}
\end{equation}
where again $\ket{\psi_i} = SWAP_{A_iA_N}\ket{\psi}$. We start by applying $\sqrt{\Pi_N}$ on the first term before then applying $\sqrt{\Pi_{N+1}}$ on the second. 

Pursuing a similar treatment to last time, we make definitions for the eigenvectors of $\Pi_N$, which are slightly cleaner this time because we do not need to differentiate between the cases $s = \frac{N+1}{2}$ and $s < \frac{N+1}{2}$. We define
\begin{align}
    \ket{\tilde{k},-,-,s,m} &= \ket{\tilde{k},s-1,s-\frac{1}{2},s,m}\\
    \ket{\tilde{k},+,-,s,m} &= \ket{\tilde{k},s+1,s+\frac{1}{2},s,m}\\
    \ket{\tilde{k},-,+,s,m} &= \sqrt{\frac{s+1}{2s+1}}\ket{\tilde{k},s,s-\frac{1}{2},s,m} + \sqrt{\frac{s}{2s+1}}\ket{\tilde{k},s,s+\frac{1}{2},s,m}\\
    \ket{\tilde{k},+,+,s,m} &= \sqrt{\frac{s}{2s+1}}\ket{\tilde{k},s,s-\frac{1}{2},s,m} - \sqrt{\frac{s+1}{2s+1}}\ket{\tilde{k},s,s+\frac{1}{2},s,m}
\end{align}
which have eigenvalues $0, 0, 0$ and $\sigma(s)$ respectively under $\Pi_N$. We once again express $\ket{\psi_i}$ in this basis and will again perform the operation
\begin{equation}
    \ket{\psi_i} = \sum_{\tilde{k},(a,b) \in \{\pm\}^2,s,m}c^{(i)}_{\tilde{k},a,b,s,m}\ket{\tilde{k},a,b,s,m} \mapsto \sum_{\tilde{k},a,b,s,m}c^{(i)}_{\tilde{k},a,b,s,m}\ket{\tilde{k}}\ket{a}\ket{b}\ket{s}\ket{m}\label{probBlockRot}
\end{equation}
(where the second and third registers on the right-hand side are each on one qubit). We describe this briefly, because it is so similar to the previous case, in fact simpler because we do not have to differentiate between $s = \frac{N+1}{2}$ and $s < \frac{N+1}{2}$.

To do the operation described in Equation \eqref{probBlockRot}, one starts by performing the inverse of the coupling stage of the Schur transform, which again is referred to simply as the Schur transform in other cases. We then perform a similar rotation to that described in Equations \eqref{detRot1} to \eqref{detRot5} on the middle three of the five registers. This will be
\begin{align}
    \ket{s-1}\ket{s-\frac{1}{2}}\ket{s} &\mapsto \ket{s-1}\ket{s-\frac{1}{2}}\ket{s}\\
    \ket{s+1}\ket{s+\frac{1}{2}}\ket{s} &\mapsto \ket{s+1}\ket{s+\frac{1}{2}}\ket{s}\\
    \ket{s}\ket{s-\frac{1}{2}}\ket{s} &\mapsto \sqrt{\frac{s+1}{2s+1}}\ket{s}\ket{s-\frac{1}{2}}\ket{s}+\sqrt{\frac{s}{2s+1}}\ket{s}\ket{s+\frac{1}{2}}\ket{s}\\
    \ket{s}\ket{s+\frac{1}{2}}\ket{s} &\mapsto \sqrt{\frac{s}{2s+1}}\ket{s}\ket{s-\frac{1}{2}}\ket{s}-\sqrt{\frac{s+1}{2s+1}}\ket{s}\ket{s+\frac{1}{2}}\ket{s}.
\end{align}
Finally, we can just compress the second and third registers into one qubit each in the same way as last time\footnote{There is one essentially unimportant detail to mention here. In this case, the entire quantum state is made up of $\ket{\psi}\ket{N+1}$ as well as $\sum_{i=1}^N\ket{\psi_i}\ket{i}$ so we can't strictly talk about registers actually changing size on just the left term. However, we can talk about qubits being added to the left-hand term when strictly they are being added to the whole expression and staying in a fixed state $\ket{0}$ on the right-hand term. Similarly, removing qubits from the left-hand term means they are returned to a fixed state $\ket{0}$ on this term and then remain untouched.}. We have therefore performed the operation of Equation \eqref{probBlockRot}.

Next, we add on the eigenvalues of $\sqrt{\Pi_N}$ in a very similar way. Reintroducing the register $\ket{r}$, which we recall is initialised in the state $\ket{0}$, we perform the following operation on the registers $\ket{a}\ket{b}\ket{s}\ket{r}$:

\begin{align}
    \ket{-}\ket{-}\ket{s}\ket{0}_r &\mapsto \ket{-}\ket{-}\ket{s}\ket{1}_r\\
    \ket{+}\ket{-}\ket{s}\ket{0}_r &\mapsto \ket{+}\ket{-}\ket{s}\ket{1}_r\\
    \ket{-}\ket{+}\ket{s}\ket{0}_r &\mapsto \ket{-}\ket{+}\ket{s}\ket{1}_r\\
    \ket{+}\ket{+}\ket{s}\ket{0}_r &\mapsto \sqrt{\frac{N}{4}\sigma(s)}\ket{+}\ket{+}\ket{s}\ket{0}_r + \sqrt{1-\left(\frac{N}{4}\sigma(s)\right)^2}\ket{+}\ket{+}\ket{s}\ket{1}_r\label{lastRot}
\end{align}
where we note that $\frac{N}{4}\sigma(s) \leq \frac{N}{N+3} < 1$. This exact choice of the coefficient $\sqrt{\frac{N}{4}\sigma(s)}$ is also important for making the algorithm for efficient, avoiding a large amount of amplitude amplification. Having attached these eigenvalues, we undo the operation of Equation \eqref{probBlockRot} so that we now have
\begin{equation}
    \frac{1}{2\sqrt{2}}\sum_{i=1}^N\sqrt{\Pi_N}\ket{\psi_i}\ket{i}\ket{0}_r + \ket{*}\ket{1}_r + \frac{1}{\sqrt{2}}\ket{\psi}\ket{N+1}.
\end{equation}
Notice that this pre-factor on the first term has appeared because of the choice of the coefficient $\sqrt{\frac{N}{4}\sigma(s)}$ in Equation \eqref{lastRot}. Applying $C-SWAP$ again gives us
\begin{equation}
    \frac{1}{2\sqrt{2}}\sum_{i=1}^N\sqrt{\Pi_i}\ket{\psi}\ket{i}\ket{0}_r + \ket{*}\ket{1}_r + \frac{1}{\sqrt{2}}\ket{\psi}\ket{N+1}.
\end{equation}
We now treat the latter term. This is much more straightforward. Writing $\ket{\psi}$ in the usual Schur basis, we may undo the Schur coupling as follows:
\begin{equation}
    \ket{\psi} = \sum_{k,j,s,m}c_{k,j,s,m}\ket{k,j,s,m} \mapsto \sum_{k,j,s,m}c_{k,j,s,m}\ket{k}\ket{j}\ket{s}\ket{m}\label{undoCoupling}.
\end{equation}
Then, reintroducing the ancilla $\ket{r}$, which again is initialised to the state $\ket{0}$, we may act on the registers $\ket{j}\ket{s}\ket{r}$ as follows (recalling that for a fixed $s$, $j$ may only take values in $\{s \pm \frac{1}{2}\}$):

\begin{align}
    \ket{s-\frac{1}{2}}\ket{s}\ket{0}_r &\mapsto \frac{1}{2}\sqrt{\tau(s)}\ket{s-\frac{1}{2}}\ket{s}\ket{0}_r + \sqrt{1-\frac{\tau(s)}{4}}\ket{s-\frac{1}{2}}\ket{s}\ket{1}_r\label{halfAdded}\\
    \ket{s+\frac{1}{2}}\ket{s}\ket{0}_r &\mapsto \ket{s+\frac{1}{2}}\ket{s}\ket{1}_r
\end{align}
noting that $\tau(s) \leq 1$, in fact achieving equality for $s = \frac{N+1}{2}$. The factor of $\frac{1}{2}$ in Equation \eqref{halfAdded} may seem superfluous at first - the reason for this will become clear very soon. Undoing the operation of Equation \eqref{undoCoupling}, we see that we have obtained overall
\begin{equation}
    \frac{1}{2\sqrt{2}}\sum_{i=1}^N\sqrt{\Pi_i}\ket{\psi}\ket{i}\ket{0}_r + \frac{1}{2\sqrt{2}}\sqrt{\Pi_{N+1}}\ket{\psi}\ket{N+1} + \ket{*}\ket{1}_r = \frac{1}{2\sqrt{2}}\sum_{i=1}^{N+1}\sqrt{\Pi_i}\ket{\psi}\ket{i}\ket{0}_r + \ket{*}\ket{1}_r\label{combining}
\end{equation}
where now the factor of $\frac{1}{2}$ in Equation \eqref{halfAdded} makes sense - it was so that there is a common pre-factor between the first and second terms of the left-hand side of Equation \eqref{combining}.

It then remains to do the amplitude amplification in exactly the same way as last time. We again define $\Pi = \ket{i=1}\bra{i=1}\otimes\ket{r=0}\ket{r=0}$ and $\tilde{\Pi} = \ket{r=0}\bra{r=0}$ and the isometry $W : \text{img}(\Pi) \to \text{img}(\tilde{\Pi})$ via $\ket{\psi}\ket{1}_i\ket{0}_r \mapsto \sum_{i=1}^{N+1}\sqrt{\Pi_i}\ket{\psi}\ket{i}\ket{0}_r$. Calling the unitary that we have constructed $U$, we see that we have
\begin{equation}
    \tilde{\Pi}U\left(\ket{\psi}\ket{i=1}\ket{r=0}\right) = \frac{1}{2\sqrt{2}}W\left(\ket{\psi}\ket{i=1}\ket{r=0}\right)
\end{equation}
and so, because we have been careful with our definitions, we only need a constant number of rounds of amplitude amplification in this case. Because $\sin\left(\frac{\pi}{10}\right) < \frac{1}{2\sqrt{2}} < \sin\left(\frac{\pi}{6}\right)$, exactly $n = 5$ rounds of amplitude amplification are required in this case. Note that in order to be precise, we must adjust the constants in Equations \eqref{lastRot} and \eqref{halfAdded} so that the unitary $U$ actually gives us
\begin{equation}
\sin\left(\frac{\pi}{10}\right)\sum_{i=1}^{N+1}\sqrt{\Pi_i}\ket{\psi}\ket{i}\ket{0}_r + \ket{*}\ket{1}_r,
\end{equation}
in Equation \eqref{combining} but this is not done above to simplify the presentation.

\subsubsection{pPBT with Maximally Entangled Resource State - Algorithm 2}\label{alternativepPBT}

We now include a further algorithm for pPBT with maximally entangled resource state, which we only include for the sake of practicality. We saw in the previous subsection that pPBT can be executed with maximally entangled resource state using only five rounds of amplitude amplification, which is quite practical. We might ask how well we can do with no amplitude amplification at all, to maximise the possibility of practical PBT for near-term devices. 

If we turn to Equation \eqref{combining}, we could start by projectively measuring the $\ket{r}$ register, obtaining the desired outcome of $r = 0$ with probability $\frac{1}{8}$. Upon such a success, one may measure the $\ket{i}$ register, succeeding in teleporting the state with the usual success probability of $\bra{\psi}\sum_{i=1}^N\Pi_i\ket{\psi}$ (and note that we get the post-measurement state that we would usually expect upon a success, which is $\sqrt{\Pi_i}\ket{\psi}$, up to normalisation). Therefore, in total we can perform pPBT with a success probability of $\frac{1}{8}$ of the usual success probability, and in particular our asymptotic (average) success probability is $\frac{1}{8}$. We will show that it is actually possible to do better than this - succeeding with asymptotic probability $\frac{1}{4}$, using no amplitude amplification. We describe this quite briefly given that there are several similarities to the previous case. 

With $\ket{\psi}$ an $(N+1)$-qubit state, we attach a $2N$-dimensional ancillary register $\ket{i}$ with orthonormal basis $\{\ket{i}\}_{i=1}^{2N}$. We will prepare a state 
\begin{equation}
    \frac{1}{2}\sum_{i=1}^N\sqrt{\Pi_i}\ket{\psi}\ket{i} + \sum_{i=N+1}^{2N}\ket{*_i}\ket{i}\label{desiredState}
\end{equation}
where the exact identities of $\ket{*_i}$ are unimportant to us. To do this, we start by creating the superposition
\begin{equation}
    \ket{\psi}\ket{1}_i \mapsto \frac{1}{\sqrt{N}}\sum_{i=1}^N\ket{\psi}\ket{i}
\end{equation}
and apply
\begin{equation}
    C-SWAP = \sum_{i=1}^NSWAP_{A_iA_N}\otimes\ket{i}\bra{i} + I \otimes \sum_{i=N+1}^{2N}\ket{i}\bra{i}
\end{equation}
to obtain
\begin{equation}
    \frac{1}{\sqrt{N}}\sum_{i=1}^N\ket{\psi_i}\ket{i}
\end{equation}
where once again $\ket{\psi_i} = SWAP_{A_iA_N}\ket{\psi}$. We now perform a very similar operation to before, writing $\ket{\psi_i}$ in terms of the eigenbasis of $\Pi_N$ and rotating to this basis:
\begin{equation}
    \ket{\psi_i} = \sum_{\tilde{k},(a,b)\in\{\pm\}^2,s,m}c^{(i)}_{\tilde{k},a,b,s,m}\ket{\tilde{k},a,b,s,m} \mapsto \sum_{\tilde{k},a,b,s,m}c^{(i)}_{\tilde{k},a,b,s,m}\ket{\tilde{k}}\ket{a}\ket{b}\ket{s}\ket{m}\label{basisRot}.
\end{equation}
In previous versions, what we've now done is attached the eigenvalues that we do want to the $\ket{r=0}$ part and rotated everything we don't want onto the $\ket{r=1}$ part. Here, we have no $\ket{r}$ register, and in fact do the following, operating on the registers $\ket{a}\ket{b}\ket{s}\ket{i}$:
\begin{align}
    \ket{-}\ket{-}\ket{s}\ket{i} &\mapsto \ket{-}\ket{-}\ket{s}\ket{N+i}\label{firstAltRot}\\
    \ket{+}\ket{-}\ket{s}\ket{i} &\mapsto \ket{+}\ket{-}\ket{s}\ket{N+i}\\
    \ket{-}\ket{+}\ket{s}\ket{i} &\mapsto \ket{-}\ket{+}\ket{s}\ket{N+i}\\
    \ket{+}\ket{+}\ket{s}\ket{i} &\mapsto \sqrt{\frac{N}{4}\sigma(s)}\ket{+}\ket{+}\ket{s}\ket{i} + \sqrt{1-\left(\frac{N}{4}\sigma(s)\right)^2}\ket{+}\ket{+}\ket{s}\ket{N+i}\label{finalRot}
\end{align}
for $i = 1, ..., N$, so that we are now putting everything we don't want onto $\{\ket{i}\}_{i=N+1}^{2N}$. Undoing the operation of Equation \eqref{basisRot} and performing C-SWAP once again leaves us with the desired state of Equation \eqref{desiredState}.

For this state, we may attempt to perform the POVM by measuring the $\ket{i}$ register projectively, obtaining an outcome in $\{1, ..., N\}$ with probability $\frac{1}{4}\bra{\psi}\sum_{i=1}^N\Pi_i\ket{\psi}$, $\frac{1}{4}$ of the usual success probability. Note also that upon such a successful measurement, one also obtains the post-measurement state normally expected from PBT: $\sqrt{\Pi_i}\ket{\psi}$, up to normalisation. It would be interesting to know if one can do better than this asymptotic (average) success probability of $\frac{1}{4}$ using similar means without any amplitude amplification or similar operations. The factor of $\frac{1}{4}$ originates from the need to have $\frac{N}{4}\sigma(s) \leq 1$ in Equation \eqref{finalRot}. Is there some deeper reason why the eigenvalue $\sigma(s)$ must be of this order and no smaller? We discuss this and similar ideas a little more in Section \ref{discussion}.

\subsubsection{pPBT with Optimised Resource State}\label{pPBTOptAlgo}

Lastly, we show how to efficiently implement the POVM associated with pPBT with optimised resource state. This will bear significant resemblance to pPBT with maximally entangled resource state. Recall that in this case, our POVM is made up of $N+1$ elements $\{\Pi_i\}_{i=1}^{N+1}$, where $\Pi_N$ has the following eigendecomposition:
\begin{alignat}{2}
&\ket{k,j,s,m} \text{ with } k_{N-2} \neq s \hspace*{0.8cm}&&\text{ have eigenvalue } 0\\
&\frac{\delta(s)}{\sqrt{\gamma(s)^2+\delta(s)^2}}\ket{\tilde{k},s,s-\frac{1}{2},s,m} - \frac{\gamma(s)}{\sqrt{\gamma(s)^2+\delta(s)^2}}\ket{\tilde{k},s,s+\frac{1}{2},s,m} \hspace*{0.8cm}&&\text{ have eigenvalue } 0\\
&\frac{\gamma(s)}{\sqrt{\gamma(s)^2+\delta(s)^2}}\ket{\tilde{k},s,s-\frac{1}{2},s,m} + \frac{\delta(s)}{\sqrt{\gamma(s)^2+\delta(s)^2}}\ket{\tilde{k},s,s+\frac{1}{2},s,m} \hspace*{0.8cm}&&\text{ have eigenvalue } u(s)\left(\gamma(s)^2+\delta(s)^2\right)
\end{alignat}
where
\begin{equation}
    \gamma(s) = \sqrt{\frac{s}{(2s+1)\nu(s-\frac{1}{2})}} \;\;\delta(s) = -\sqrt\frac{s+1}{(2s+1)\nu(s+\frac{1}{2})} \;\;u(s) = \frac{2^{N+1}h(N)(2s+1)}{Ng^{[N-1]}(s)}
\end{equation}
and
\begin{equation}
    \nu(j) = \frac{2^Nh(N)(2j+1)}{g^{[N]}(j)} \;\; g^{[N]}(j) = \frac{(2j+1)N!}{\left(\frac{N}{2}-j\right)!\left(\frac{N}{2}+1+j\right)!} \;\; h(N) = \frac{6}{(N+1)(N+2)(N+3)}
\end{equation}
and once again we have that $\Pi_i = SWAP_{A_iA_N}\Pi_NSWAP_{A_iA_N}$ for $i = 1, ..., N-1$. The eigendecomposition for $\Pi_{N+1}$ may then be easily stated. The operator has an eigenbasis $\ket{k,j,s,m}$ with corresponding eigenvalues
\begin{equation}
    1-\frac{2^{N-1}\lambda(j,s)u(s)}{\nu(j)}
\;\;\text{ where }\;\;
\lambda(j,s) = \begin{cases} \frac{1}{2^N}\left(\frac{N}{2}-j\right) & \text{ if } s = j + \frac{1}{2} \\ \frac{1}{2^N}\left(\frac{N}{2} + j + 1\right) & \text{ if } s = j - \frac{1}{2}\end{cases}.
\end{equation}
We use a very similar strategy to pPBT with maximally entangled resource state, except this time there is no benefit to be gained from the superposition of Equation \eqref{superposition} because too many of the eigenvalues are of constant magnitude in this case. We therefore simply start by creating the superposition
\begin{equation}
    \ket{\psi}\ket{1}_i \mapsto \frac{1}{\sqrt{N+1}}\sum_{i=1}^{N+1}\ket{\psi}\ket{i}
\end{equation}
where $\ket{i}$ is an ancillary register of dimension $N+1$ and $\ket{\psi}$ is an $(N+1)$-qubit state. We act with $C-SWAP$ as in Equation \eqref{CSWAP} to get
\begin{equation}
    \frac{1}{\sqrt{N+1}}\sum_{i=1}^N\ket{\psi_i}\ket{i} + \frac{1}{\sqrt{N+1}}\ket{\psi}\ket{N+1}
\end{equation}
where, as before, $\ket{\psi_i} = SWAP_{A_iA_N}\ket{\psi}$. We then make the expected definitions for the eigenbasis of $\Pi_N$:

\begin{align}
    \ket{\tilde{k},-,-,s,m} &= \ket{\tilde{k},s-1,s-\frac{1}{2},s,m}\\
    \ket{\tilde{k},+,-,s,m} &= \ket{\tilde{k},s+1,s+\frac{1}{2},s,m}\\
    \ket{\tilde{k},-,+,s,m} &= \frac{\delta(s)}{\sqrt{\gamma(s)^2+\delta(s)^2}}\ket{\tilde{k},s,s-\frac{1}{2},s,m} - \frac{\gamma(s)}{\sqrt{\gamma(s)^2+\delta(s)^2}}\ket{\tilde{k},s,s+\frac{1}{2},s,m}\\
    \ket{\tilde{k},+,+,s,m} &= \frac{\gamma(s)}{\sqrt{\gamma(s)^2+\delta(s)^2}}\ket{\tilde{k},s,s-\frac{1}{2},s,m} + \frac{\delta(s)}{\sqrt{\gamma(s)^2+\delta(s)^2}}\ket{\tilde{k},s,s+\frac{1}{2},s,m}
\end{align}
and, writing $\ket{\psi_i}$ in terms of this eigenbasis, we rotate to it using the usual prescription of inverting the Schur coupling and performing a rotation and compression on the $\ket{a}\ket{b}\ket{s}$ registers:
\begin{equation}
    \ket{\psi_i} = \sum_{\tilde{k},(a,b)\in \{\pm\}^2,s,m} c^{(i)}_{\tilde{k},a,b,s,m}\ket{\tilde{k},a,b,s,m} \mapsto \sum_{\tilde{k},(a,b)\in \{\pm\}^2,s,m} c^{(i)}_{\tilde{k},a,b,s,m}\ket{\tilde{k}}\ket{a}\ket{b}\ket{s}\ket{m}\label{newBlockRot}
\end{equation}
We use, as always, one qubit in an ancillary register $\ket{r}$ to perform the block-encoding. We act on the registers $\ket{a}\ket{b}\ket{s}\ket{r}$ in the expected fashion:
\begin{align}
    \ket{-}\ket{-}\ket{s}\ket{0}_r &\mapsto \ket{-}\ket{-}\ket{s}\ket{1}_r\\
    \ket{+}\ket{-}\ket{s}\ket{0}_r &\mapsto \ket{+}\ket{-}\ket{s}\ket{1}_r\\
    \ket{-}\ket{+}\ket{s}\ket{0}_r &\mapsto \ket{-}\ket{+}\ket{s}\ket{1}_r\\
    \ket{+}\ket{+}\ket{s}\ket{0}_r &\mapsto \sqrt{u(s)\left(\gamma(s)^2+\delta(s)^2\right)}\ket{+}\ket{+}\ket{s}\ket{0}_r + \sqrt{1-u(s)^2\left(\gamma(s)^2+\delta(s)^2\right)^2}\ket{+}\ket{+}\ket{s}\ket{1}_r
\end{align}
where we are guaranteed that $u(s)\left(\gamma(s)^2+\delta(s)^2\right) \leq 1$ by virtue of the fact that $\{\Pi_i\}_{i=1}^{N+1}$ is a valid POVM and so $\Pi_N \leq I$. Undoing the operation of Equation \eqref{newBlockRot} before applying $C-SWAP$ again gives us
\begin{equation}
    \frac{1}{\sqrt{N+1}}\sum_{i=1}^N\sqrt{\Pi_i}\ket{\psi}\ket{i}\ket{0}_r + \ket{*}\ket{1}_r + \frac{1}{\sqrt{N+1}}\ket{\psi}\ket{N+1}
\end{equation}
where the identity of $\ket{*}$ is unimportant to us. We then treat $\sqrt{\Pi_{N+1}}$ in a similar way to in the case of pPBT with maximally entangled resource state. We may write $\ket{\psi}$ in terms of the usual Schur basis and undo the Schur coupling:
\begin{equation}
    \ket{\psi} = \sum_{k,j,s,m}c_{k,j,s,m}\ket{k,j,s,m} \mapsto \sum_{k,j,s,m}c_{k,j,s,m}\ket{k}\ket{j}\ket{s}\ket{m}\label{schurCouplingLast}
\end{equation}
before acting on $\ket{j}\ket{s}\ket{r}$ to attach the eigenvalues as usual:
\begin{equation}
    \ket{j}\ket{s}\ket{0}_r \mapsto \sqrt{1-\frac{2^{N-1}\lambda(j,s)u(s)}{\nu(j)}}\ket{j}\ket{s}\ket{0}_r + \sqrt{1-\left(1-\frac{2^{N-1}\lambda(j,s)u(s)}{\nu(j)}\right)^2}\ket{j}\ket{s}\ket{1}_r.
\end{equation}
where we are again guaranteed that $0\leq1-\frac{2^{N-1}\lambda(j,s)u(s)}{\nu(j)}\leq1$ given that $0\leq \Pi_{N+1}\leq 1$ since $\{\Pi_i\}_{i=1}^{N+1}$ is a valid POVM. By then undoing the operation of Equation \eqref{schurCouplingLast}, we get
\begin{equation}
    \frac{1}{\sqrt{N+1}}\sum_{i=1}^{N+1}\sqrt{\Pi_i}\ket{\psi}\ket{i}\ket{0}_r + \ket{*}\ket{1}_r
\end{equation}
which we may then amplify. With the exact same treatment as in the case of dPBT, we may amplify the $r=0$ eigenspace to create the desired state $\sum_{i=1}^{N+1}\sqrt{\Pi_i}\ket{\psi}\ket{i}\ket{0}_r$ using $n = \mathcal{O}(\sqrt{N})$ rounds of oblivious amplitude amplification.

\section{Discussion}\label{discussion}

In this work, we presented efficient quantum algorithms for the optimal POVMs for the four regimes of port-based teleportation defined in Ishizaka and Hiroshima's work \cite{ishizaka2009quantum}. We present explicit gate complexities and ancilla counts for each protocol. One very interesting question surrounds the given gate complexities, where we have seen that these bounds vary between the paradigms due to the amount of amplitude amplification required. We saw that pPBT with maximally entangled resource state was implemented more efficiently than the other two cases arising from the lower need for amplitude amplification. This came from the use of the superposition of Equation \eqref{superposition} and the rotation of Equation \eqref{lastRot} and this, in turn, was possible because of the magnitude of the eigenvalues of the POVM operators in this case being of the order of $\frac{1}{N}$ (for all but $\Pi_{N+1}$), as opposed to the other regimes where some of the eigenvalues of all the POVM operators are of constant order. It seems, therefore, that greater efficiency is possible with these methods in these cases where the POVM operators have small eigenvalues.

It is known that the optimisation problems relevant to PBT can have multiple solutions - indeed dPBT with optimised resource state does have multiple optimising POVMs and resource states, as discussed. Morally, the optimisation problem does not care about gate complexity or efficiency generally. Therefore, it might be interesting to view the relevant optimisation problems from a complexity theoretic standpoint and see if alternative optimising solutions exist that allow for more efficient algorithms.

Even further, one could consider non-optimal solutions that make up for their lack of optimality with greater efficiency - this is in the spirit of our alternative algorithm for pPBT with maximally entangled resource state (Section \ref{alternativepPBT}) which required no amplitude amplification at all, but had asymptotic success probability only $\frac{1}{4}$.

Finally, we note that one problem that remains open is the efficient implementation of multi-port-based teleportation \cite{kopszak2021multiport,mozrzymas2021optimal,studzinski2022efficient}, which allows for the teleportation of composite systems with greater efficiency than repeated, or `packaged' ordinary PBT. Whether this can be described via the usual Schur-Weyl duality relied on in this work, or requires `twisted' Schur-Weyl duality \cite{fei2023efficient,grinko2023gelfand,nguyen2023mixed}, is unknown.

\vspace*{1cm}

\noindent \textit{Acknowledgements}: The authors wish to thank Dmitry Grinko and M\=aris Ozols for informative and lengthy discussions both on their PBT algorithms and the differing notions of Schur transforms that exist. Thanks is also extended to Felix Leditzky for discussions on the optimality of the pretty good measurement for dPBT. SS acknowledges support from the Royal Society University Research Fellowship and ``Quantum Simulation Algorithms for Quantum Chromodynamics'' grant (ST/W006251/1).

\bibliography{references}

\appendix

\section{Complexity Analysis for the PBT Algorithms}\label{complexityAnalysis}

We will now provide the complexity analysis for the algorithms. We recall that we provide four algorithms in total, for each of which there are two possible versions. These four algorithms are for dPBT in Section \ref{dPBTAlgo}, for pPBT with maximally entangled resource state - 2 different algorithms are presented in Sections \ref{pPBTMESAlgo} and \ref{alternativepPBT} - and lastly for pPBT with optimised resource state in Section \ref{pPBTOptAlgo}. The two versions of each algorithm arise because one has an option over whether to use the Schur transform as in \cite{bacon2005quantum} (the BCH Schur transform) or \cite{wills2023generalised} (the spin coupling Schur transform). The former runs faster - with gate complexity $n\text{poly}(\log(n),\log(1/\epsilon))$ for accuracy $\epsilon$, but uses more ancillas - $\mathcal{O}(n\log(n))$. In contrast, the algorithm of \cite{wills2023generalised} runs with gate complexity $\mathcal{O}(n^3\log(n)\log(n/\epsilon))$ but uses $\mathcal{O}(\log(n))$ ancillas.

To obtain explicit gate complexities for these algorithms in terms of the Clifford + T universal set, we use the same technique as employed in \cite{kirby2017practical}. As in \cite{kirby2017practical}, we use the fact that an arbitrary unitary $U$ may be decomposed into a sequence of two-level rotations whose length equals the number of non-zero elements on or below the diagonal of $U$, not counting 1's on the diagonal\footnote{It is typical to upper bound this by the number of non-zero elements in $U$, not counting 1's on the diagonal.}. Subsequently, a two-level rotation on $n$ qubits may be decomposed to an accuracy of $\delta$ in $\mathcal{O}(n\log(1/\delta))$ Clifford + T gates. This latter fact allows us to translate between a `two-level rotation complexity' i.e. the number of two-level rotations in a given sequence of two-level rotations, and `Clifford + T complexity'.

We note that in one round of amplitude amplification, the complexity of implementing $C_\Pi NOT$ and $C_{\tilde{\Pi}} NOT$ (for our particular $\Pi$ and $\tilde{\Pi}$) and single qubit gates (which are special cases of two-level rotations) is sub-leading compared to that of the unitary we implement in each block, $U$. It can also be noted that every two-level rotation we employ throughout each algorithm acts only on $\mathcal{O}(\log(N))$ qubits, since every operation throughout the algorithm acts only on $\mathcal{O}(\log(N))$ qubits. As such, suppose that the two-level rotation gate complexity for $U$, without the complexity of the Schur transform, is $p$. Each of these $p$ two-level rotations should then be decomposed to accuracy $\mathcal{O}\left(\frac{\epsilon}{pn}\right)$ in the Clifford + T gate set, where $n$ is the number of rounds of amplitude amplification used, which can be done in $\mathcal{O}\left(\log(N)\log(pn/\epsilon)\right)$ such gates. The whole unitary $U$ in one round of amplitude amplification then has a Clifford + T gate complexity $T_{Sch}$ + $p\log(N)\log\left(\frac{pn}{\epsilon}\right)$, where $T_{Sch}$ is the Clifford + T gate complexity of whichever Schur coupling we are using, the BCH Schur transform \cite{bacon2005quantum}, or the spin coupling transform \cite{wills2023generalised}, implemented to accuracy $\mathcal{O}\left(\frac{\epsilon}{n}\right)$. We therefore obtain a Clifford + T gate complexity for the entire algorithm of $\mathcal{O}\left[\left(T_{Sch} + p\log(N)\log\left(\frac{pn}{\epsilon}\right)\right)n\right]$ with an overall accuracy of $\epsilon$ for the Naimark unitary.

We have the following values of $p$ and $n$ for each algorithm.

\begin{table}[h]
    \centering
    \begin{tabular}{|c|c|c|c|c|}
         \hline& dPBT & \makecell{pPBT\\Maximally Entangled Resource State\\Algorithm 1}& \makecell{pPBT\\Maximally Entangled Resource State\\Algorithm 2} & \makecell{pPBT\\Optimised Resource State\\\hspace*{1cm}}  \\\hline\hline
         $p$& $\mathcal{O}(N)$&$\mathcal{O}(N)$&$\mathcal{O}(N)$&$\mathcal{O}(N)$\\
         $n$ & $\mathcal{O}(\sqrt{N})$ & $\Theta(1)$&$\Theta(1)$&$\mathcal{O}(\sqrt{N})$\\\hline
         
    \end{tabular}
    \caption{The values of $p$ (the two-level rotation gate complexity for each round of amplitude amplification, without the Schur coupling) and $n$ (the number of rounds of amplitude amplification).}
    \label{pAndnForEach}
\end{table}
Further, we have the values for $T_{Sch}$:

\begin{equation}
    T_{Sch} = \begin{cases} N \text{poly}(\log(N),\log(1/\epsilon)) & \text{BCH Schur Transform \cite{bacon2005quantum}}\\ \mathcal{O}(N^3\log(N)\log(N/\epsilon)) &\text{Spin Coupling Schur Transform \cite{wills2023generalised}}\end{cases}
\end{equation}
where we recall that we differentiate the two transforms because the former uses more ancillary qubits $(\mathcal{O}(N\log(N))$ than the latter $(\mathcal{O}(\log(N))$.

It would be somewhat tedious to go through and prove each value of $p$ individually. Instead, we prove the value for dPBT in some detail and then provide two points that are important to mention for why we get the same value of $p$ for all the pPBT algorithms - otherwise the arguments are essentially exactly the same for each value of $p$ for each algorithm.

For dPBT, we analyse the relevant operations as follows within the unitary $U$ that do not include the Schur coupling:

\begin{itemize}
    \item The initial superposition $\ket{1}_i \mapsto \frac{1}{\sqrt{N}}\sum_{i=1}^N\ket{i}$, may be implemented via, for example, the quantum Fourier transform, $QFT_N$, which has gate complexity $\text{polylog}(N)$.
    
    \item The operation $C-SWAP = \sum_{i=1}^NSWAP_{A_iA_N}\otimes \ket{i}\bra{i}$ may be re-written as 
    
    $\Pi_{i=1}^N\left(SWAP_{A_iA_N}\otimes\ket{i}\bra{i} + I \otimes \sum_{j \neq i}\ket{j}\bra{j}\right)$ i.e. a product of $N$ operators where each operator swaps $A_i$ and $A_N$ controlled on the ancilla being in the state $\ket{i}$. Each of these $N$ operators has exactly two non-zero elements (other than 1's on the diagonal): $\ket{0}_{A_i}\ket{1}_{A_N}\ket{i} \leftrightarrow \ket{1}_{A_i}\ket{0}_{A_N}\ket{i}$.

    \item After the Schur coupling, we perform the rotation given by Equations \eqref{detRot1} to \eqref{detRot5}. Note that this is already defined as a unitary on the whole space and it has $\mathcal{O}(N)$ non-zero entries. The reason for this is that $s$ may take $\mathcal{O}(N)$ values, and so all three registers may take $\mathcal{O}(N)$ values, due to the fact that adjacent spins may differ from each other only by $\pm \frac{1}{2}$. We see that each resulting superposition is over two computational basis elements, so each column in the operation's matrix has weight at most two.

    \item In the same way, the compression step of Equations \eqref{compressionDet1} to \eqref{compressionDet2} has two-level rotation gate complexity $\mathcal{O}(N)$.

    \item Then, when the eigenvalues are attached in Equations \eqref{eigenvalueAttachDet1} to \eqref{nonTrivRot}, this operation can easily be extended to a unitary on the whole space:
    \begin{align}
        \ket{-}\ket{-}\ket{s}\ket{0}_r &\leftrightarrow \ket{-}\ket{-}\ket{s}\ket{1}_r\\
        \ket{+}\ket{-}\ket{s}\ket{0}_r &\leftrightarrow \ket{+}\ket{-}\ket{s}\ket{1}_r\\
        \ket{-}\ket{+}\ket{s}\ket{0}_r &\leftrightarrow \ket{-}\ket{+}\ket{s}\ket{1}_r\\
        \ket{+}\ket{+}\ket{s}\ket{0}_r &\mapsto \sqrt{\alpha(s)^2+\beta(s)^2}\ket{+}\ket{+}\ket{s}\ket{0}_r + \sqrt{1-(\alpha(s)^2+\beta(s)^2)}\ket{+}\ket{+}\ket{s}\ket{1}_r\\
        \ket{+}\ket{+}\ket{s}\ket{1}_r &\mapsto \sqrt{1-(\alpha(s)^2+\beta(s)^2)}\ket{+}\ket{+}\ket{s}\ket{0}_r - \sqrt{\alpha(s)^2+\beta(s)^2}\ket{+}\ket{+}\ket{s}\ket{1}_r
    \end{align}
    Again, this has two-level rotation gate complexity $\mathcal{O}(N)$ because $\ket{s}$ takes $\mathcal{O}(N)$ values and each resulting superposition is over only two elements.
\end{itemize}

We may therefore conclude $p = \mathcal{O}(N)$ for dPBT. We now explain why for the second algorithm of pPBT with maximally entangled resource state we may obtain $p = \mathcal{O}(N)$ as well. In particular, the operation described by Equations \eqref{firstAltRot} to \eqref{finalRot}, repeated below, looks less efficient than this:

\begin{align}
    \ket{-}\ket{-}\ket{s}\ket{i} &\mapsto \ket{-}\ket{-}\ket{s}\ket{N+i}\label{copyRot1}\\
    \ket{+}\ket{-}\ket{s}\ket{i} &\mapsto \ket{+}\ket{-}\ket{s}\ket{N+i}\\
    \ket{-}\ket{+}\ket{s}\ket{i} &\mapsto \ket{-}\ket{+}\ket{s}\ket{N+i}\\
    \ket{+}\ket{+}\ket{s}\ket{i} &\mapsto \sqrt{\frac{N}{4}\sigma(s)}\ket{+}\ket{+}\ket{s}\ket{i} + \sqrt{1-\left(\frac{N}{4}\sigma(s)\right)^2}\ket{+}\ket{+}\ket{s}\ket{N+i}\label{copyRot4}.
\end{align}
Since the registers $\ket{s}$ and $\ket{i}$ can both take $\mathcal{O}(N)$ values, it appears initially that this operation has two-level rotation gate complexity of $\mathcal{O}(N^2)$. However, to make this more efficient, one may just flip a single qubit in the ancillary register $\ket{i}$ to map $\ket{i}\mapsto\ket{N+i}$. More explicitly, the $2N$-dimensional register $\ket{i}$ may be encoded into $\ket{\iota}\ket{y}$, where $\ket{\iota}$ is an $N$-dimensional register, and $\ket{y}$ is a single qubit. We encode $\ket{i = 1, ..., N}$ into $\ket{\iota = 1, ..., N}\ket{y = 0}$ and $\ket{i=N+1, ..., 2N}$ into $\ket{\iota = 1, ..., N}\ket{y=1}$. Then the operation described by Equations \eqref{copyRot1} to \eqref{copyRot4} does not, in fact, act on $\ket{\iota}$, only acting on $\ket{\pm}\ket{\pm}\ket{s}\ket{y}$, and so we get a desirable $\mathcal{O}(N)$ two-level rotation gate complexity here.

We make one more point as to why we may get $p = \mathcal{O}(N)$ for the pPBT algorithms, after which the values of $p$ follow in the exact same way as for dPBT. In particular, this point relates to algorithm 1 of pPBT with maximally entangled resource state as well as pPBT with optimised resource state. In both of these cases the ancillary register $\ket{i}$ had dimension $N+1$ and there are several operations that we wish to control on this register. In particular, in both algorithms, the unitary $U$ within each amplitude amplification round can be summarised by the two steps
\begin{enumerate}
    \item $\sum_{i=1}^N\ket{\psi}\ket{i}\ket{0}_r\mapsto\sum_{i=1}^N\sqrt{\Pi_i}\ket{\psi}\ket{i}\ket{0}_r + \ket{*}\ket{1}_r$
    \item $\ket{\psi}\ket{N+1}\ket{0}_r \mapsto \sqrt{\Pi_{N+1}}\ket{\psi}\ket{N+1}\ket{0}_r + \ket{*}\ket{1}_r$
\end{enumerate}
so that we are acting on $\ket{i=1, ..., N}$ in the first step and $\ket{i=N+1}$ in the second step. Again, at first this makes it appear that we lose efficiency because we are seemingly acting on the whole $\ket{i}$ register in addition to our other steps. However, in a very similar way to the point just made - we can encode the register $\ket{i}$ into some $\ket{\iota}\ket{y}$, where $\ket{\iota}$ is $N$-dimensional and $\ket{y}$ is a single qubit. $\ket{i=1, ..., N}$ is then encoded into $\ket{\iota = 1, ..., N}\ket{y=0}$ and $\ket{i=N+1}$ may be encoded into $\ket{\iota = 1}\ket{y=1}$, so that we may simply control our operations on the one qubit $\ket{y}$, thus not losing any efficiency.

\section{Different Notions of Schur Transforms and the Necessity for a Pre-Mapping Stage in some Applications}\label{preMapping}

We now discuss the two primary notions of the quantum Schur transform that exist, aiming to clear up some confusion that arises between the two of them. In particular, we will show how the traditional view on the Schur transform is insufficient for certain applications in quantum information, and that a ``pre-mapping stage'' as in \cite{wills2023generalised} (or something similar) is essential in these cases. We present these definitions on qubits only - the extension to qudits is natural - see \cite{alex2011numerical} for more details on the representation theory of $SU(d)$.

In the work on Permutational Quantum Computing (PQC) in \cite{jordan2009permutational}, the Schur basis is considered to be a simultaneous spin eigenbasis of $n$ commuting, complete observables. With $C^{J,M}_{j_1, m_1; j_2, m_2}$ the usual $SU(2)$ Clebsch-Gordan coefficients, a Schur state may be written as

\begin{equation}
    \ket{j_1, j_2, ..., j_{n-2}, J, M} = \sum_{x \in \{\pm\frac{1}{2}\}^n} C^{j_1, m_1}_{\frac{1}{2},x_1;\frac{1}{2},x_2}C^{j_2,m_2}_{j_1, m_1; \frac{1}{2}, x_3}...C^{j_{n-2},m_{n-2}}_{j_{n-3},m_{n-3};\frac{1}{2},x_{n-1}}C^{J,M}_{j_{n-2},m_{n-2};\frac{1}{2},x_n}\ket{x_1...x_n}\label{schurStateAppendix}
\end{equation}
where $m_i$ is used as shorthand for $x_1 + ... + x_{i+1}$. Note that here we are writing computational basis states as $\ket{\frac{1}{2}}$ and $\ket{-\frac{1}{2}}$ as opposed to $\ket{0}$ and $\ket{1}$, just for the sake of convenience. It must be emphasised that these states are simply $n$-qubit states.

These states are labelled by a sequence of spin eigenvalues $j_1, ..., j_{n-2}, J$ and $M$ where $j_i \in \frac{\mathbb{N}_0}{2}$, $J \in \frac{\mathbb{N}_0}{2}$ and $M \in \{-J, -J+1, ..., J-1, J\}$. Furthermore, each consecutive element of $j_1, ..., j_{n-2}, J$ must differ from each other by either $+\frac{1}{2}$ or $-\frac{1}{2}$, and we note that these values may not fall below zero. Because the Schur basis forms a basis for $(\mathbb{C}^2)^{\otimes n}$, there must necessarily be $2^n$ valid such sequences $j_1, ..., j_{n-2}, J, M$.

In the work on PQC \cite{jordan2009permutational}, the Schur transform is defined as the \textit{unitary} operation that maps a computational basis state $\ket{x_1...x_n}$ to a Schur state $\ket{j_1, ..., j_{n-2},J,M}$ coherently over all computational basis states. To which Schur state each $\ket{x_1...x_n}$ gets mapped is left as an arbitrary choice. In PQC, one starts with some fixed $n$-qubit computational basis state $\ket{x}$, performs the Schur transform to produce some fixed Schur state, performs a permutation of the qubits, before inverting the Schur transform and measuring in the computational basis\footnote{In the original paper \cite{jordan2009permutational}, more general spin eigenbases were considered than the Schur basis (or sequentially coupled basis) alone. These are, however, nothing more than bases instantiating the decomposition of Schur-Weyl duality with different subgroup adaptations.}. PQC itself was later shown to be classically simulable \cite{SchurSampling,havlicek2018quantum}, however `upgraded' versions, dubbed PQC+, may be considered for which the qubit permutation is replaced by some more complicated gate, see for example \cite{zheng2022super} (believed to be non-classical and to have practical applications).

We now consider an alternative definition of the Schur transform as in \cite{harrow2005applications,kirby2017practical}. Here, the Schur transform is taken to be an operation that performs the mapping

\begin{equation}
    \ket{j_1, j_2, ..., j_{n-2}, J, M} \mapsto \ket{j_1}\ket{j_2}...\ket{j_{n-2}}\ket{J}\ket{M}\label{mathematicalSchurTransform}
\end{equation}
coherently over all Schur states $\ket{j_1, j_2, ..., j_{n-2}, J, M}$, which are exactly as defined in Equation \eqref{schurStateAppendix}. The right-hand side of Equation \eqref{mathematicalSchurTransform} is a computational basis state on \textit{more than $n$ qubits}. Each register $\ket{j_i}$, $\ket{J}$ and $\ket{M}$ is formed of exactly as many qubits as is needed to encode the given eigenvalue. For example, one can check that $j_3$ may take values in the set $\{0, 1, 2\}$, and so two qubits make up this register. Because each eigenvalue takes at most linearly many values, each register may be formed of $\mathcal{O}(\log(n))$ qubits, meaning that the right-hand side of Equation \eqref{mathematicalSchurTransform} is a computational basis state on $\mathcal{O}(n\log(n))$ qubits. In this case, the Schur transform has therefore been defined as an \textit{isometry} rather than a unitary operation, since we have performed a mapping into a larger Hilbert space.

We call the first notion of the Schur transform, defined by a unitary mapping a computational basis state to a Schur state the `physical' notion of the Schur transform, and the notion of the Schur transform mapping each Schur basis state to its encoding on computational basis states as in Equation \ref{mathematicalSchurTransform} the `mathematical' notion of the Schur transform. We justify this characterisation with the fact that the origins of permutational quantum computing may be traced back to spin networks \cite{marzuoli2005computing, penrose1971angular}, whereas the latter notion has a closer relation to Schur-Weyl duality.

There are two primary differences between the notions of the quantum Schur transform that may each lead to confusion. First, they are in a non-precise sense inverses of each other. They are not actually inverses of each other - one is a unitary and one is an isometry. However, the physical Schur transform is mapping $n$-qubit computational basis states to $n$-qubit Schur states, whereas the mathematical Schur transform is mapping $n$-qubit Schur states to computational basis states (on $\mathcal{O}(n\log(n))$ qubits). One may not need to necessarily see this as a change in definition, one may choose to instead see this as a difference in transformation convention, where the physical Schur transform gives an `active' transformation convention, and the mathematical Schur transform uses a `passive' transformation convention. 

The second crucial difference between the two is that the first is a unitary operation, whereas the second is an isometry, mapping into a larger space with entangled ancillary qubits, as explained. Some efficient implementations of the Schur transform are only implementing the mathematical transform \cite{harrow2005applications,kirby2017practical}, whereas others can perform both \cite{wills2023generalised,krovi2019efficient}. These latter algorithms that can perform the unitary operation are said to be performing a `clean' Schur transform. 

A smaller difference between the physical and mathematical perspective is notational, which should be easily surmountable. Indeed, in the context of Schur-Weyl duality, Schur states $\ket{j_1, ..., j_{n-2},J,M}$ may be denoted as, for example, $\ket{\lambda,q_\lambda,p_\lambda}$, whereas their encodings in a computational basis state $\ket{j_1}\ket{j_2}...\ket{j_{n-2}}\ket{J}\ket{M}$ may be denoted as $\ket{\lambda}\ket{q_\lambda}\ket{p_\lambda}$, as in \cite{bacon2005quantum}, for example. Here, the total angular momentum/spin of all the qubits, $J$, corresponds to $\lambda$, the partition of $n$, whereas $q_\lambda$, which indexes within irreps of the unitary group, corresponds to $M$, and $p_\lambda$, the symmetric group irrep index, corresponds to all of $(j_1, j_2, ..., j_{n-2})$.

We will now discuss the algorithm of \cite{wills2023generalised} briefly, which we might refer to as the spin coupling transform, as we feel that it is the most simple of the available algorithms for a clean, unitary quantum Schur transform.

The algorithm takes place in two stages. First is the so-called `pre-mapping' stage, in which one maps

\begin{equation}
    \ket{x_1...x_n} \mapsto \ket{j_1}\ket{j_2}...\ket{j_{n-2}}\ket{J}\ket{M}\label{preMappingAppendixEqn}
\end{equation}
coherently over all $n$-qubit computational basis states $x$. Ancillary qubits have therefore been introduced in this operation\footnote{Note there is a version of this algorithm for which only $\mathcal{O}(\log(n))$ qubits are required here, rather than $\mathcal{O}(n\log(n))$, but we do not go into this for the sake of brevity.}. As mentioned, the choice of to which $(j_1, ..., j_{n-2},J,M)$ each $x$ is mapped is arbitrary - the user may have a particular choice, such as that given by the RSK correspondence \cite{krattenthaler2006growth}, but we leave this choice as arbitrary here. It is worth appreciating the fact that the mapping of Equation \eqref{preMappingAppendixEqn}, despite being a map between computational basis states, is non-trivial to implement efficiently. Only particular computational basis states on the right-hand side define valid eigenvalue sequences $\ket{j_1}\ket{j_2}...\ket{j_{n-2}}\ket{J}\ket{M}$ and one must view all the qubits in such a computational basis state to make sure that it corresponds to such a valid eigenvalue sequence. Thus how to perform this pre-mapping efficiently is not immediately obvious.

The second stage of the algorithm of \cite{wills2023generalised} is the `coupling stage', performing 

\begin{equation}
    \ket{j_1}\ket{j_2}...\ket{j_{n-2}}\ket{J}\ket{M} \mapsto \ket{j_1, j_2, ..., j_{n-2}, J, M}.\label{couplingStageAppendixEqn}
\end{equation}
As such, the mathematical Schur transform is exactly the inverse of the coupling stage of the spin coupling transform. It is worth noting that in the present paper, for the PBT algorithms, only the inverse of the coupling stage (which many people know as simply the Schur transform) is ever needed. Indeed, for many applications, this operation is sufficient. It is, however, important for some applications that the full, clean, unitary can be performed, and our primary intention with this appendix is to clean up this point for algorithm designers, hoping to make clear the necessity of a pre-mapping stage (or just a clean Schur transform more generally) in certain applications.

Indeed, consider a situation where one wishes to perform the inverse Schur transform (as a unitary), perform some arbitrary $n$-qubit unitary $U$, and then perform the forwards Schur transform. Recalling that there are $2^n$ valid encodings $\ket{j_1}\ket{j_2}...\ket{j_{n-2}}\ket{J}\ket{M}$, just as there are $2^n$ computational basis states $\ket{x}$, it is tempting to attempt to perform $U$ directly on the $\ket{j_1}\ket{j_2}...\ket{j_{n-2}}\ket{J}\ket{M}$, by only acting on the $2^n$ levels that form valid encodings. It is not obvious, however, how to do this efficiently for a general $n$-qubit unitary operation $U$. One has to check every qubit of $\ket{j_1}\ket{j_2}...\ket{j_{n-2}}\ket{J}\ket{M}$ in order to determine if it is a valid encoding of the eigenvalues. Thus, $U$ cannot be implemented efficiently (without further information) on these registers, more specifically because the $2^n$ valid encodings do not have a tensor product structure.

As such, these types of algorithms, or similarly an algorithm like PQC+ \cite{zheng2022super}, where the forwards unitary transform is used, followed by some unitary, followed by the inverse unitary transform, do necessitate a `clean' transform like \cite{wills2023generalised} or \cite{krovi2019efficient}.

\end{document}